\documentclass[pdflatex,sn-mathphys,a4paper]{sn-jnl}
\usepackage{amsmath,commath,graphicx,url,hyperref,microtype,subcaption}
\usepackage{latexsym,courier,upgreek,xcolor,ulem,enumitem}
\hypersetup{colorlinks,citecolor=blue,linkcolor=red,urlcolor=blue}
\usepackage[displaymath]{lineno}
\begin{document}

\title[Compressibility in two-phase flow]{Steady-state two-phase flow
  of compressible and incompressible fluids in a capillary tube of
  varying radius}

\author[1,2]{\fnm{Hyejeong} \sur{L. Cheon}}
\author[2]{\fnm{Hursanay} \sur{Fyhn}}
\author[2]{\fnm{Alex} \sur{Hansen}}
\author[1]{\fnm{{\O}ivind} \sur{Wilhelmsen}}
\author*[3]{\fnm{Santanu} \sur{Sinha}}\email{santanu.sinha@ntnu.no}

\affil[1]{
  \orgdiv{PoreLab, Department of Chemistry},
  \orgname{Norwegian University of Science and Technology (NTNU)},
  \orgaddress{\city{Trondheim}, \postcode{N-7491}, \country{Norway}}}

\affil[2]{
  \orgdiv{PoreLab, Department of Physics},
  \orgname{Norwegian University of Science and Technology (NTNU)},
  \orgaddress{\city{Trondheim}, \postcode{N-7491}, \country{Norway}}}

\affil*[3]{
  \orgdiv{PoreLab, Department of Physics},
  \orgname{University of Oslo (UiO)},
  \orgaddress{\city{Oslo}, \postcode{N-0371}, \country{Norway}}}

\abstract{We study immiscible two-phase flow of a compressible and an
  incompressible fluid inside a capillary tube of varying radius under
  steady-state conditions. The incompressible fluid is Newtonian and
  the compressible fluid is an inviscid ideal gas. The surface tension
  associated with the interfaces between the two fluids introduces
  capillary forces that vary along the tube due to the variation in
  the tube radius. The interplay between effects due to the capillary
  forces and the compressibility results in a set of properties that
  are different from incompressible two-phase flow. As the fluids move
  towards the outlet, the bubbles of the compressible fluid grows in
  volume due to the decrease in pressure. The volumetric growth of the
  compressible bubbles makes the volumetric flow rate at the outlet
  higher than at the inlet. The growth is not only a function of the
  pressure drop across the tube, but also of the ambient
  pressure. Furthermore, the capillary forces create an effective
  threshold below which there is no flow. Above the threshold, the
  system shows a weak non-linearity between the flow rates and the
  effective pressure drop, where the non-linearity also depends on the
  absolute pressures across the tube.}


\maketitle

\section{Introduction}
Hydrodynamic properties of the flow of multiple immiscible and
incompressible fluids, otherwise known as two-phase flow
\cite{b88,d92,b17,ffh22}, are controlled by a number of different
factors: fluid properties such as the viscosity contrast and surface
tension between the fluids, driving parameters such as the applied
pressure drop or the flow rate, and geometrical properties of the
medium such as the size and shape of the pore space. The combined
effects of these factors make two-phase flow different and more
complex than single phase flow. The dimensionless parameters that play
a key role to define the flow properties are the ratio between the
viscous and capillary forces, referred to as the capillary number, and
the ratio between the viscosities of the two fluids. Depending on the
values of these parameters, the flow generates different types of
fingering patterns \cite{cw85, lz85, mfj85, lmt04, zmp19} or stable
displacement fronts \cite{ltz88} during invasion processes where one
fluid displaces another in the porous medium.

Displacement processes are transient. If one continues to inject after
breakthrough, the flow enters a steady state characterized by a
situation where the macroscopic flow properties fluctuate or remain
constant around well-defined averages. A more general form of
steady-state flow can be achieved by continuously injecting both
fluids simultaneously. In this case, the dynamics at the pore scale
might have fluid clusters breaking up and forming, while the
macroscopic flow parameters still have well-defined averages.

Over the last decade, it has become clear that steady-state flow
deviates from the linear Darcy relationship \cite{d56} between the
total flow rate and pressure drop over a range of parameters. Rather,
one finds a power law relationship between pressure drop and the
volumetric flow rate \cite{tkr09,rcs11} in that range. In terms of the
capillary number, this range is intermediary, with linearity appearing
both for lower and higher values \cite{sbd17, glb20,
  zbg21}. Theoretical work to understand the physics behind the
non-linearity has appeared in e.g.\ \cite{tlk09, sh12, zbg21}, and
computational studies have been performed using Lattice Boltzmann
simulations \cite{yts13} and dynamic pore network modeling
\cite{sgv21, sbd17}. It is now believed that a fundamental mechanism
behind this non-linearity is the capillary barriers at the pore
throats, which create an effective yield threshold. When the viscous
forces increase, they overcome the capillary barriers creating new
flow paths. This increases the effective mobility and thus the
non-linear behavior appears \cite{rh87}. The disorder in the
pore-space properties, such as the pore-size distribution \cite{rsh21}
and the wetting angle distribution \cite{fsr21}, therefore play key
roles in determining the value of the exponent relating the volumetric
flow rate and the pressure drop in the non-linear regime.

The majority of the analytical and numerical approaches mentioned
above consider the two fluids to be incompressible, whereas many of
the experiments and applications use air as one of the fluids. Air is
strongly compressible, which introduces complex pore-scale mechanisms
such as trapping and coalescence \cite{l41, ly94}. Compressibility is
relevant to a wide range of applications with liquid and gas transport
in porous media, for example, ${\rm CO}_2$ transport and storage
\cite{rk15, lkd15, ipr19} and the transport in fuel cells
\cite{nmn20}. Another class of applications where the compressibility
plays a key role are those involving phase transitions of the fluids
such as boiling and condensation. There are industrial applications
where such processes are of high importance, for example aerospace
vehicle thermal protection \cite{hzl17}, high power electronics
cooling systems \cite{gzk11, lww12, lwy20} and chemical reactors
\cite{bs19}. These applications utilize the high specific surface area
of a porous medium with fluid flowing inside, which enhances the heat
and mass transfer rates \cite{sgd16, szx11}. There are also natural
processes such as drying of soil \cite {rn94} where a liquid to gas
transition takes place.

In this paper we present a study of two-phase flow of a mixture of
compressible and incompressible fluids in a capillary tube with
varying radius. We consider two fluids, one is an incompressible
Newtonian fluid obeying Poiseuille flow whereas the other is a
compressible ideal gas, where the viscosity is assumed to be
negligible. The fluids flow as a series of bubbles and droplets under
a constant pressure drop along the tube.

In case of two-phase flow of two incompressible fluids in a
corresponding capillary tube, it has been found that the volumetric
flow rate depends on the square root of the pressure drop along the
tube minus a threshold pressure \cite{shb13}. The primary goal of the
present work is to determine how this constitutive equation changes
when one of the two fluids is compressible.

A secondary goal of this work is to provide a basis for dynamic pore
network modeling \cite{b01, mt09, jh12, sgv21} of
compressible-incompressible fluid mixtures. This opens the possibility
for incorporating thermodynamic effects in such models such as
boiling.  We note that the other dominating computational model in
this context, the Lattice Boltzmann model \cite{grz91, rin12}, can
only incorporate fluids that are weakly compressible \cite{qyz17,
  gfj20}.

We describe in Section \ref{govern} the equations that govern the flow
through the capillary tube. In Section \ref{boundaries} we introduce
the boundary conditions used, i.e., how we inject alternate
compressible and incompressible fluid into the tube. Section
\ref{updating} describes how the governing equations are integrated in
time.

Section \ref{results} presents the results of our
investigation. Section \ref{steady} defines what we mean by
steady-state flow in the context of expanding bubbles. In section
\ref{bubble} we investigate how the compressible bubbles grow as they
advance along the tube, thus increasing the overall flow rate of the
fluids. Section \ref{effective} presents the relation between
volumetric flow rate and pressure drop at both the inlet and outlet.

We summarize our results in Section \ref{conclusion}. Section
\ref{suppl} contains the description of the videos provided in the
electronic supplementary material.

\section{Methodology}
\label{methodology}

\begin{figure}[htbp]
  \centerline{\includegraphics[width=0.8\textwidth]{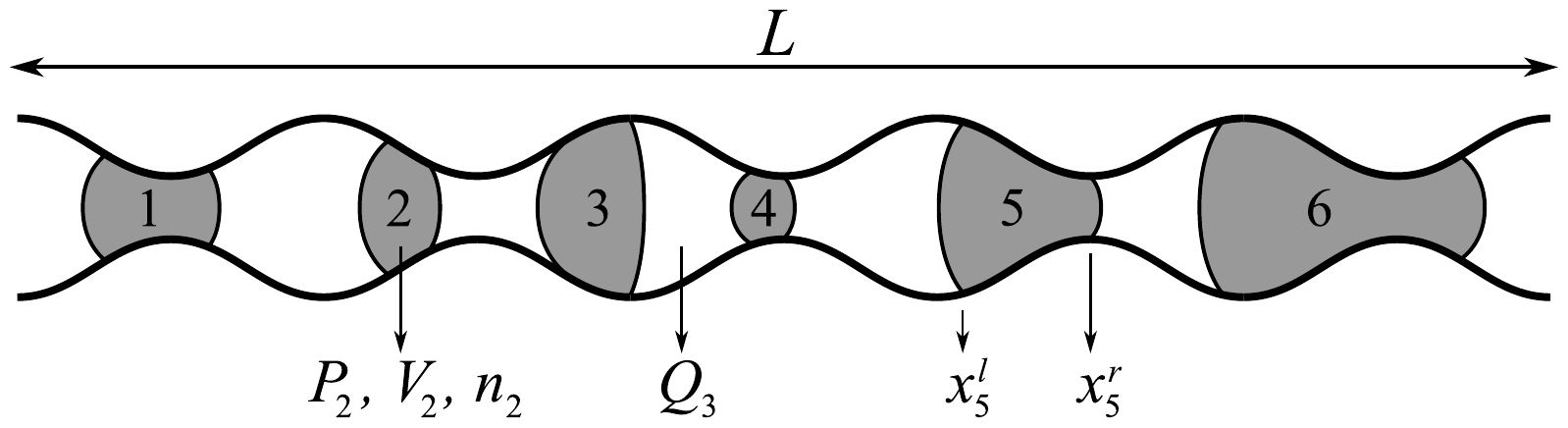}}
  \caption{\label{fig_tubedemo}Illustration of the tube geometry and
    the indexed variables. The shaded fluid represents the non-wetting
    compressible fluid and the white fluid represents the more wetting
    incompressible liquid. There are $N=6$ bubbles here indicated
    by the numbers $i=1,\ldots ,6$. The indexed variables $P_i$, $V_i$
    and $n_i$ respectively correspond to the pressure, volume and
    moles of the $i$th bubble whereas $Q_i$ corresponds to the
    flow rate of the droplet between $i$th and $(i+1)$th
    bubbles.}
\end{figure}

The capillary tube considered in this work is filled with an
incompressible and a compressible fluid, immiscible to each other,
which flow through it. The radius of the capillary tube varies along
the flow direction, $x$. The fluids are separated by menisci,
generating a surface tension. The incompressible fluid is a viscous
Newtonian liquid obeying Hagen-Poiseuille flow whereas the
compressible fluid is an inviscid ideal gas. The flow occurs as a plug
flow with a series of alternate bubbles and droplets of the two fluids
as illustrated in Figure \ref{fig_tubedemo}. There is no fluid film
along the tube walls and therefore no coalescence or snap off taking
place inside the tube during the flow. We will refer the compressible
and the incompressible fluid segments as {\it bubbles} and {\it
  droplets} respectively.

\subsection{Governing equations}
\label{govern} 

We assume that at a given time the system contains $N$ compressible
bubbles denoted by $i=1,2,\ldots,N$ from left to right as shown in
Figure \ref{fig_tubedemo}. The volume $V_i$ and the pressure $P_i$ of
the $i$th bubble are connected through the ideal gas law,
\begin{equation}
  \label{eqn_pcfl}
  \displaystyle
  P_iV_i = n_iRT \;,
\end{equation}
where $n_i$ is the number of moles of gas present inside the bubble,
$R$ is the ideal gas constant and $T$ is the temperature. The tube is
assumed to have a constant average cross-sectional area ($A$) in terms
of the fluid volume and therefore $V_i=A(x_i^r-x_i^l)$ where $x_i^l$
and $x_i^r$ are the positions of the left and right menisci of the
$i$th bubble respectively.

The volume of an incompressible droplet on the other hand will remain
constant throughout the flow and the flow rate will depend on the
pressures of the two compressible bubbles bordering it. The volumetric
flow rate of the incompressible droplet between $i$ and $i+1$ is
denoted by $Q_i$, and follows the constitutive equation \cite{w21},
\begin{linenomath}
  \begin{equation}
  \label{eqn_qifl}
  \displaystyle
  Q_i =  \frac{A^2}{8\pi\mu(x_{i+1}^l-x_{i}^r)}\left[P_i - P_{\rm c}(x_i^r) - P_{i+1}  + P_{\rm c}(x_{i+1}^l)\right] \;,
  \end{equation}
\end{linenomath}
where $\mu$ is the viscosity of the incompressible fluid and $P_c(x)$
is the capillary pressure at $x$. Here we consider the incompressible
fluid to be more wetting with respect to the pore walls than the
compressible fluid, thus determining the sign of $P_c$ in Equation
\ref{eqn_qifl}. We model $P_c$ by using the Young-Laplace equation
\cite{d92},
\begin{equation}
  \label{eqn_pc}
  \displaystyle
   P_{\rm{c}}(x) = \frac{2\gamma}{r(x)} \;,
\end{equation}
where $r(x)$ is the radius of the tube at $x$. Here $\gamma =
\sigma\cos(\theta)$ where $\sigma$ is the surface tension between the
fluids and $\theta$ is the wetting angle of the fluid with respect to
the tube wall. The variation in the radius of the tube shown in Figure
\ref{fig_tubedemo} is modeled by
\begin{equation}
  \label{eqn_tube}
  \displaystyle
  r(x) = \frac{1}{2}\left[w + 2a\cos\left(\frac{2h\pi x}{L}\right)\right]
\end{equation}
where $L$ is the tube length, $w$ is the average radius, $a$ is the
amplitude of oscillation and $h$ is the number of periods.

\subsection{Boundary conditions}
\label{boundaries}

\begin{figure}[ht]
  \centerline{\hfill\includegraphics[width=\textwidth]{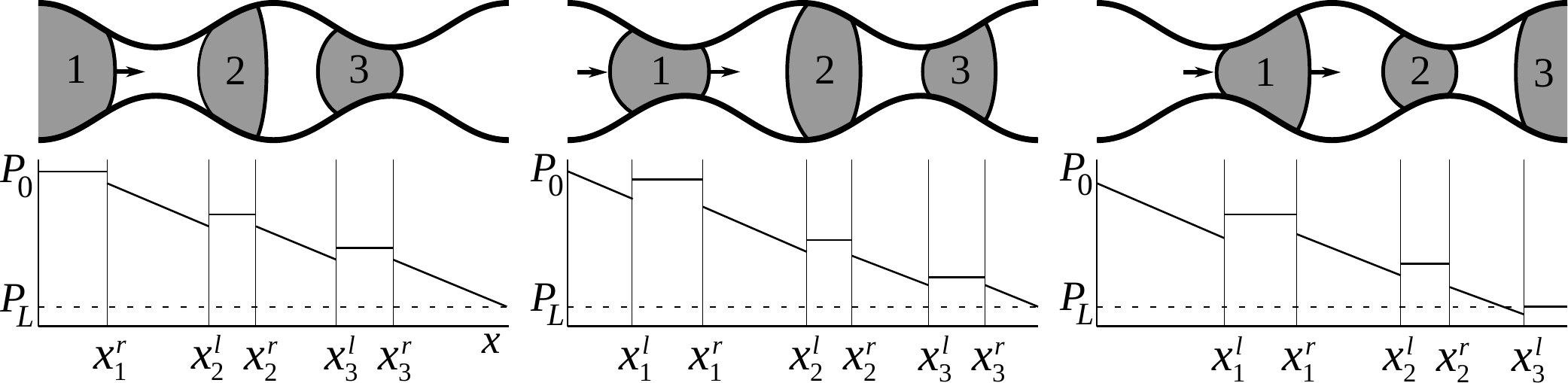}\hfill}
  \centerline{\hfill (a) \hfill \hfill (b) \hfill \hfill (c) \hfill}
  \caption{\label{fig_configs}Illustration of different configurations
    where bubbles and droplets are colored as gray and white
    respectively. In (a), a bubble is entering at the inlet and
    therefore $P_1 = P_0$ there. In (c), a bubble is leaving at the
    outlet, therefore $P_N=P_L$ there. A droplet is entering at the
    inlet in (b) and (c), and leaving at the outlet in (a) and
    (b). The flow rates of such droplets are calculated using
    Equation \ref{eqn_qinout}.}
\end{figure}

The system is driven by a constant pressure drop $\Delta P = P_0-P_L$
where $P_0$ and $P_L$ are the pressures at the inlet ($x=0$) and
outlet ($x=L$) respectively. The two fluids are injected alternatively
at the inlet. Depending on which fluid that is being injected and
which fluid that is leaving the tube, there will be different
configurations as illustrated in Figure \ref{fig_configs}. When a
bubble is entering at the inlet [Figure \ref{fig_configs}(a)] or
leaving at the outlet [Figure \ref{fig_configs}(c)], the pressure in
that bubble is given by $P_0$ or $P_L$ respectively. This is because
the compressible fluid has no viscosity and thus the pressure inside a
bubble is uniform. The pressures inside all other bubbles are
calculated using Equation \ref{eqn_pcfl}. When a droplet is entering
at the inlet [Figure \ref{fig_configs}(b) and (c)] or leaving at the
outlet [Figure \ref{fig_configs}(a) and (b)], the respective flow
rates $Q_0$ and $Q_N$ are given by,
\begin{equation}
  \label{eqn_qinout}
  \begin{split}
    \displaystyle
    Q_0 & = \frac{A^2}{8\pi\mu x_1^l}\left[P_0 - P_1  + P_{\rm c}(x_1^l)\right] \; {\rm and}\\
    Q_N & = \frac{A^2}{8\pi\mu(L-x_N^r)}\left[P_N - P_{\rm c}(x_N^r) - P_L\right] \;,
  \end{split}
\end{equation}
whereas the flow rates of the remaining droplets are calculated using
Equation \ref{eqn_qifl}.

The simulation is started with the tube completely filled with the
incompressible fluid. The two fluids are then injected alternately
through the inlet using small time steps. Whenever the injection is
switched to a different fluid, a new menisci is created and the
injection is continued for that fluid until the bubble or the droplet
being injected has reached a given length, $b_{\rm C}$ or $b_{\rm I}$
respectively. For each new bubble or droplet, a new value for $b_{\rm
  C}$ or $b_{\rm I}$ is determined using the following scheme:
\begin{equation}
  \label{eqn_bsize}
  \displaystyle
  b_{\rm C} = b_{\rm min} + kF_{\rm C}b_{\rm max} \quad \; {\rm and} \quad b_{\rm I} = b_{\rm min} + kF_{\rm I}b_{\rm max} \;,
\end{equation}
where $k$ is chosen from a uniform distribution of random numbers
between $0$ and $1$. $F_{\rm C}$ and $F_{\rm I}$ are the tentative
values of the fractional flows for the bubbles and droplets
respectively. The two parameters $b_{\rm max}$ and $b_{\rm min}$ set
the smallest and largest allowed sizes of any bubble or droplet. We
consider here $b_{\rm min}=L/10^4$ and $b_{\rm max}=L/50$. The
parameters $b_{\rm C}$ and $b_{\rm I}$ decide the initial sizes of the
bubbles and droplets just after they detach from the inlet. For the
compressible fluid, this determines the number of moles $n_i$ inside a
bubble,
\begin{equation}
  \label{eqn_molecfl}
  \displaystyle
  n_i = \frac{Ab_{\rm C}P_0}{RT} \;,
\end{equation}
which remains constant for that bubble throughout the flow after it
gets detached from the inlet.

\subsection{Updating the menisci}
\label{updating}

At any time, the two menisci bordering a droplet inside the tube move
with the same velocities. The velocities of the menisci are calculated
from the velocities $v_i$ of the droplets using Equations
\ref{eqn_qifl} and \ref{eqn_qinout},
\begin{equation}
  \label{eqn_velo}
  \displaystyle
  \od{{x_{i}^r}}{t} = \od{{x_{i+1}^l}}{t} = v_i = \frac{Q_i}{A} \;.
\end{equation}
We solve these ordinary differential equations using an explicit Euler
scheme, thus updating positions of all menisci by choosing a small
time step $\Delta t$.

Depending on the position of the menisci and the corresponding
capillary pressures, the bubbles may compress or expand. If a bubble
compresses at any time step, it means the left and right interfaces of
that bubble approach each other. This necessitates the choice of time
step $\Delta t$ to be sufficiently small, as otherwise, the two
menisci around that bubble will collapse after the time step. We deal
with this situation in the following way. First we calculate a time
$\Delta t_1$ that is needed to pass one pore-volume of incompressible
fluid through the tube,
\begin{equation}
  \label{eqn_delt1}
  \displaystyle
  \Delta t_1 = \frac{8\pi\mu L^2}{A(P_0-P_L)} \;.
\end{equation}
Next, we check for every bubble $i$ if $(v_{i-1}-v_i)>0$, that is,
whether the two menisci bordering the bubble are approaching each
other in that time step. If this criterion is found to be true for any
of the bubbles $j$, we measure the time it will take for the two
menisci to collapse,
\begin{equation}
  \label{eqn_delt2}
  \displaystyle
  \Delta t_2^j = \frac{x_j^r-x_j^l}{v_{j-1}-v_j} \;.
\end{equation}
After calculating $\Delta t_1$ and $\Delta t_2^j$, we determine a time
$\Delta t$ for that step from,
\begin{equation}
  \label{eqn_delt}
  \displaystyle
  \Delta t = {\rm min}(a^*\Delta t_1, b^*\Delta t_2^j) \;,
\end{equation}
which means that if there is a possibility for a bubble to collapse
during the time step, we chose $\Delta t$ from the minimum of
$a^*\Delta t_1$ and all of $b^*\Delta t_2^j$. If there is no
possibility of collapse, we use $\Delta t$ equal to $a^*\Delta
t_1$. For the simulations presented in this paper, we set
$a^*=10^{-8}$ and $b^*=10^{-6}$.

\section{Results and discussions}
\label{results}

We perform steady-state simulations considering a tube of length
$L=100\,{\rm cm}$ with $w=1\,{\rm cm}$, $a=0.25\,{\rm cm}$ and $h=30$
(Equation \ref{eqn_tube}). The viscosity of the incompressible fluid
is $\mu=0.001\,{\rm Pa.s}$, the ideal gas constant is $R=8.31\,{\rm
  J/(mol.K)}$ and the temperature is kept fixed throughout the
simulation at $T=293\,{\rm K}$. We fix $F_c=0.4$ (Equation
\ref{eqn_bsize}) which sets the volumetric fractional flow of the
compressible fluid at the inlet around that value. We perform
simulations varying the pressure drops ($\Delta P = P_0-P_L$) as well
as the absolute outlet pressure with different values of the surface
tension, $\gamma$.

\begin{figure}[htbp]
  \centerline{\hfill\includegraphics[width=0.5\textwidth]{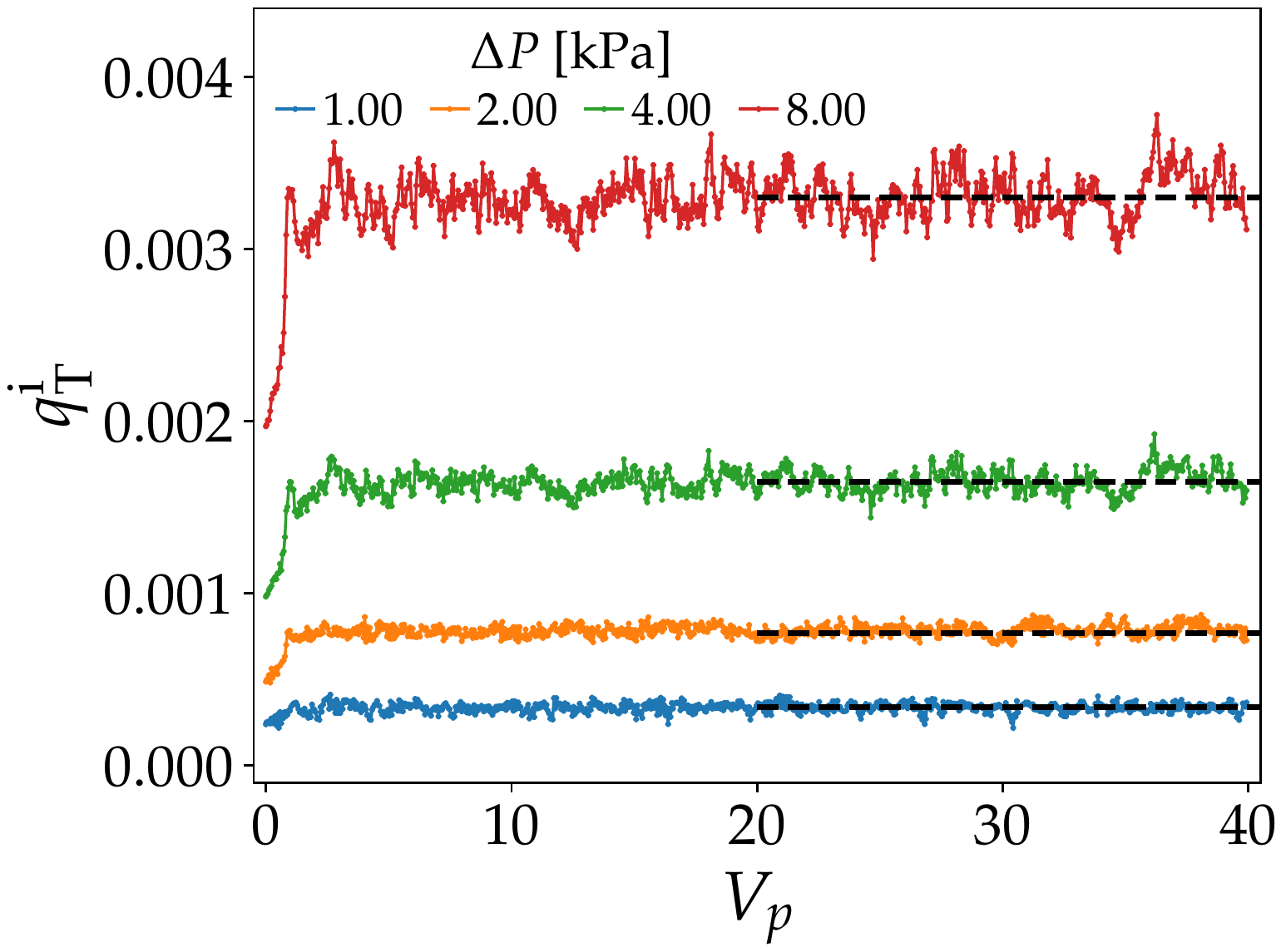}\hfill}
  \caption{\label{fig_qttime}Total volumetric flow rate $q_{\rm
      T}^{\rm i}$ at the inlet as a function of the injected pore
    volume $V_p$ for the outlet pressure $P_L=1\,{\rm kPa}$ and the
    surface tension $\gamma=0.09\,{\rm N/m}$ and for the pressure
    drops $\Delta P=1,2,4$ and $8\,{\rm kPa}$ respectively. The
    steady-state values of the flow rates are measured by taking
    averages in the range of $20$ to $40$ pore volumes as indicated by
    the dashed lines.}
\end{figure}

\subsection{Steady-state flow}
\label{steady}

The steady state is defined by the volumetric flow rates of the fluids
fluctuating around a stable average. Due to the expansion of the
compressible fluid, which we will discuss in a moment, the volumetric
flow rate of the fluids changes as the fluids flow towards the
outlet. We define the quantities $Q_{\rm T}^{\rm i}$, $Q_{\rm C}^{\rm
  i}$, $Q_{\rm I}^{\rm i}$ as the average steady-state flow rates for
the total, compressible and incompressible fluids at the inlet and
$Q_{\rm T}^{\rm o}$, $Q_{\rm C}^{\rm o}$, $Q_{\rm I}^{\rm o}$ as those
at the outlet. The inlet and outlet flow rates are measured by
tracking the displacement of the first meniscus nearest to the inlet
and the last meniscus near the outlet, which are either the left or
the right meniscus of the first ($i=1$) and the last ($i=N$)
bubbles. The instantaneous flow rates of the bubbles and droplets are
measured as $q_{\rm C}^{\rm i}=A\sum\Delta x_1^r/\sum \Delta t$ for
$x_1^l=0$, $q_{\rm I}^{\rm i}=A\sum\Delta x_1^l/\sum \Delta t$ for
$x_1^l>0$ and $q_{\rm C}^{\rm o}=A\sum\Delta x_N^l/\sum \Delta t$ for
$x_N^l=L$, $q_{\rm I}^{\rm o}=A\sum\Delta x_N^r/\sum \Delta t$ for
$x_N^R<L$. This measurement is performed after every $0.05$
pore-volumes of fluid are injected and the sum is therefore over the
time steps in between. The total flow rates are therefore given by,
$q_{\rm T}^{\rm i,o} = q_{\rm C}^{\rm i,o}+q_{\rm I}^{\rm i,o}$. This
provides the measurement of the injected and outlet flow rates as a
function of the injected pore volumes or of the time. In Figure
\ref{fig_qttime}, we plot $q_{\rm T}^{\rm i}$ as a function of the
pore-volumes ($V_p$) injected for $P_L=1\,{\rm kPa}$. The pore-volume
$V_p$ is defined as the ratio between the volume of the inject fluids
and the volume of the total pore space of the tube, which provides an
estimate of how many times the pore space was flushed with the
fluids. The plots show that $q_{\rm T}^{\rm i}$ increases with time at
the beginning of the flow. This increase in $q_{\rm T}^{\rm i}$ is due
to the decrease in the effective viscosity of the system caused by the
injection of inviscid compressible gas into the tube filled with
viscous incompressible fluid. After the injection of a few pore
volumes, $q_{\rm T}^{\rm i}$ fluctuates around a constant average
($Q_{\rm T}^{\rm i}$) shown by the horizontal dashed lines which
defines the steady state. We run our simulations for $40$ pore volumes
of fluid where the steady-state averages are taken after $20$ pore
volumes injected to ensure that a steady state has been reached.

\subsection{Bubble growth}
\label{bubble}

As a compressible bubble moves along the tube, the volume of the
bubble increases due to the decrease in the pressure towards the
outlet \cite{vls10}. The bubble can also grow due to other mechanisms,
such as the increase in temperature or a phase transition between
liquid and gas phases \cite{w98,kwd06}, but these phenomena are not
studied here. A simulation with a single bubble inside a short tube is
shown in the supplementary material which illustrates that the bubble
increases in size as it flows towards the outlet. To understand how
this growth depends on different flow parameters in the steady state,
we define the growth function $G_{\rm C}(x)$ by,
\begin{equation}
  \label{eqn_gcdef}
  \displaystyle
  G_{\rm C}(x) = \frac{V(x) - V_{\rm 0}}{V_0}\;,
\end{equation}
where $V_0$ and $V(x)$ are the volume of a given bubble initially
after detaching from the inlet and when its center is at $x$. We
measure $G_{\rm C}$ by including all the bubbles that are not attached
to the inlet or outlet and calculate the time average value of
$(V(x)-V_0)/V_0$ in the investigated time interval, where $x$ is the
center of the bubble.

\begin{figure}[htbp]
  \centerline{\hfill
    \includegraphics[width=0.45\textwidth]{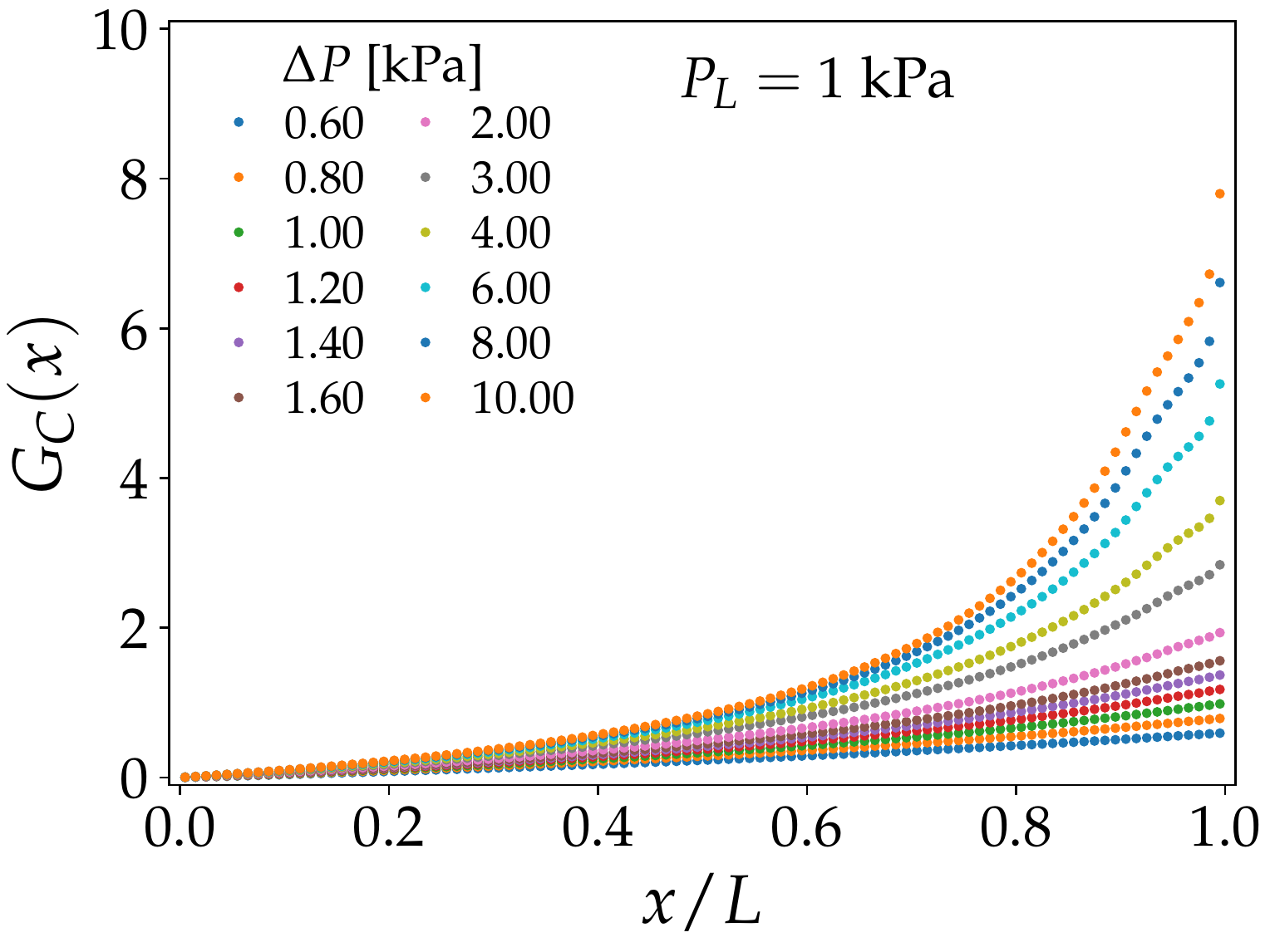}\hfill
    \includegraphics[width=0.45\textwidth]{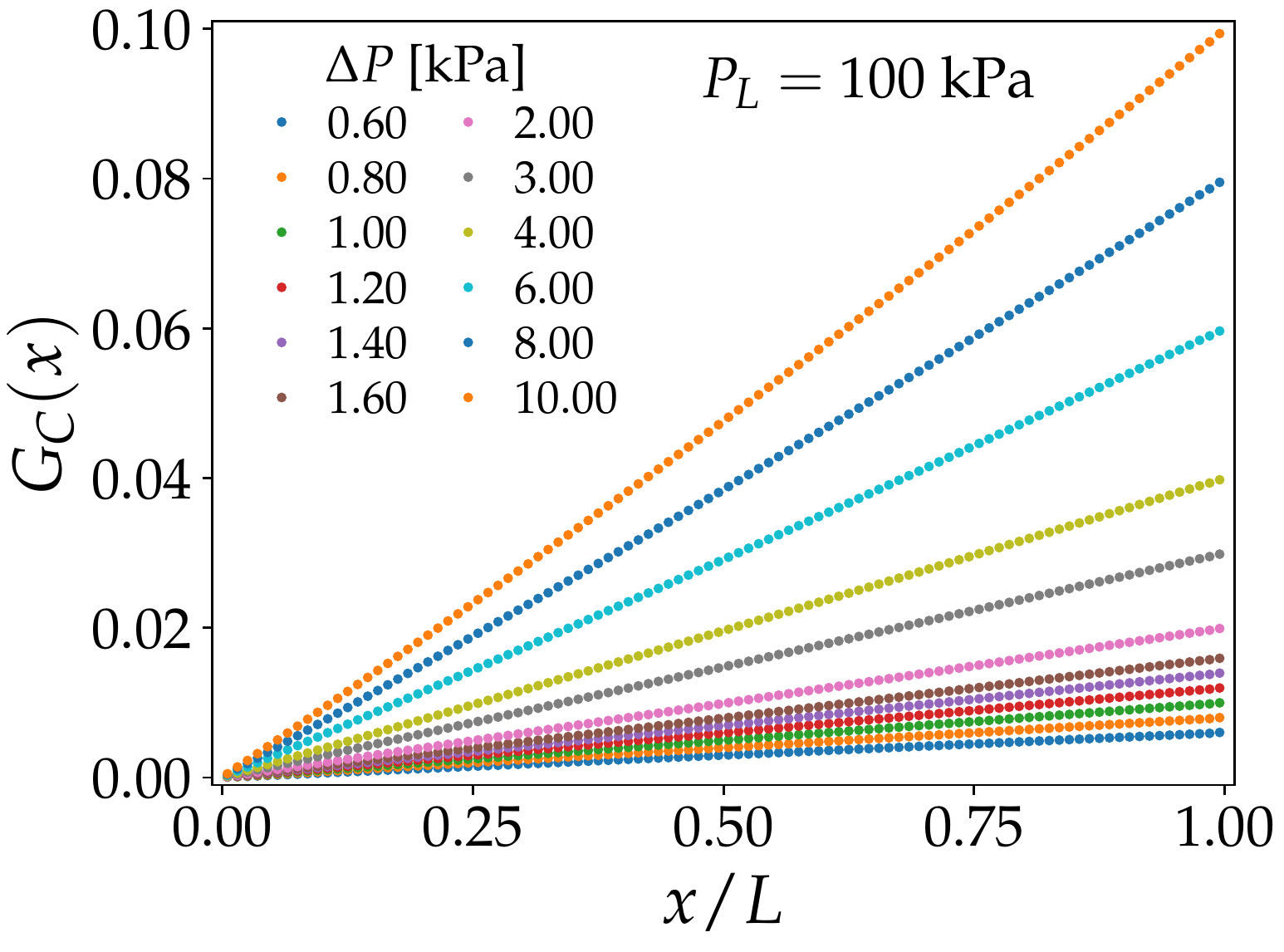}\hfill}
  \centerline{\hfill (a) \hfill \hfill (b) \hfill}
  \caption{\label{fig_gcuscl}Plot of the bubble growth $G_{\rm C}(x)$
    in the steady state as a function of the scaled position $x/L$
    inside the tube for zero surface tension, $\gamma = 0$. The two
    plots show the results for the same set of pressure drops $\Delta
    P$ with different outlet pressures $P_L$.}
\end{figure}

\begin{figure}[htbp]
  \centerline{\hfill
    \includegraphics[width=0.45\textwidth]{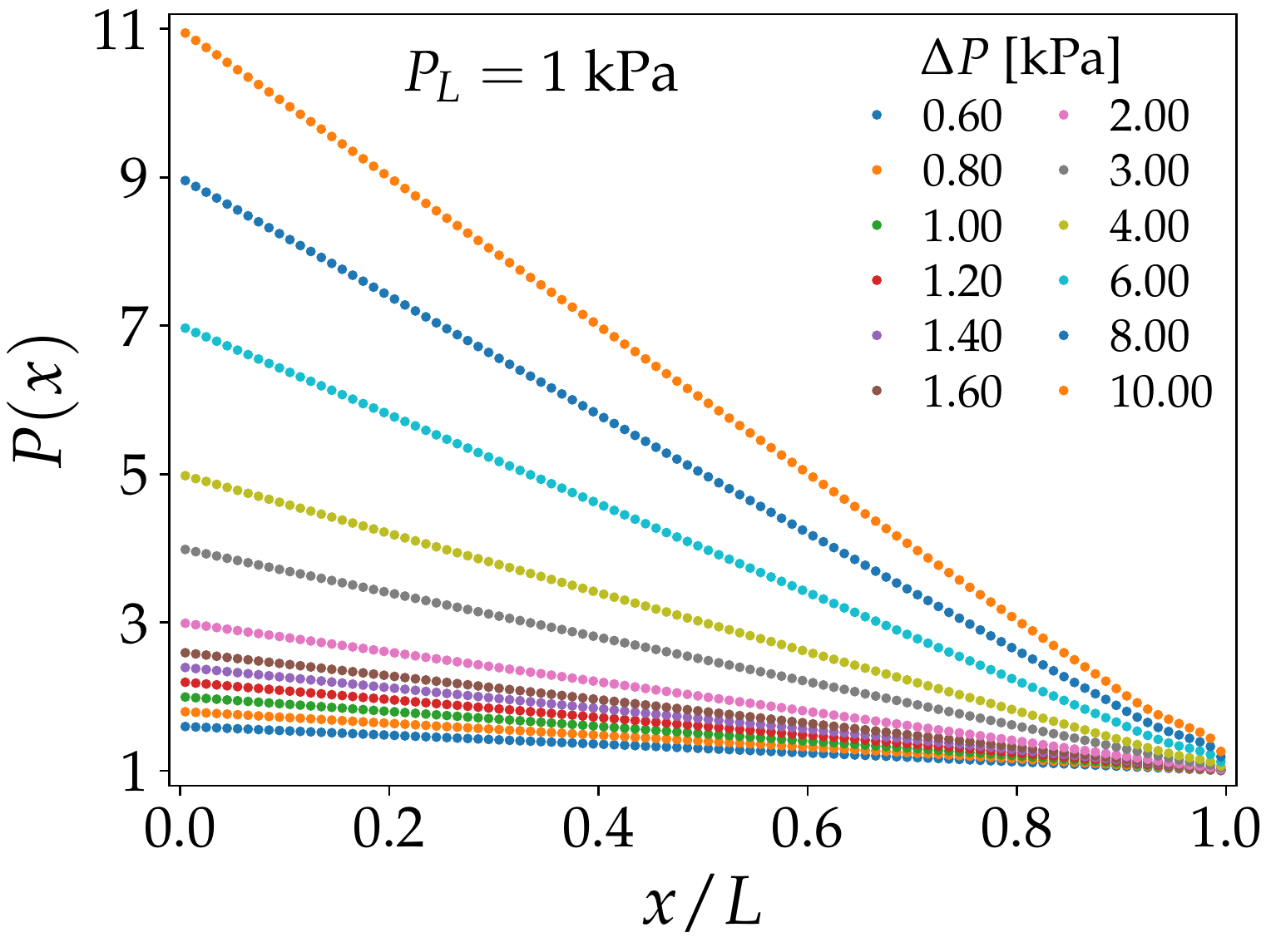}\hfill
    \includegraphics[width=0.45\textwidth]{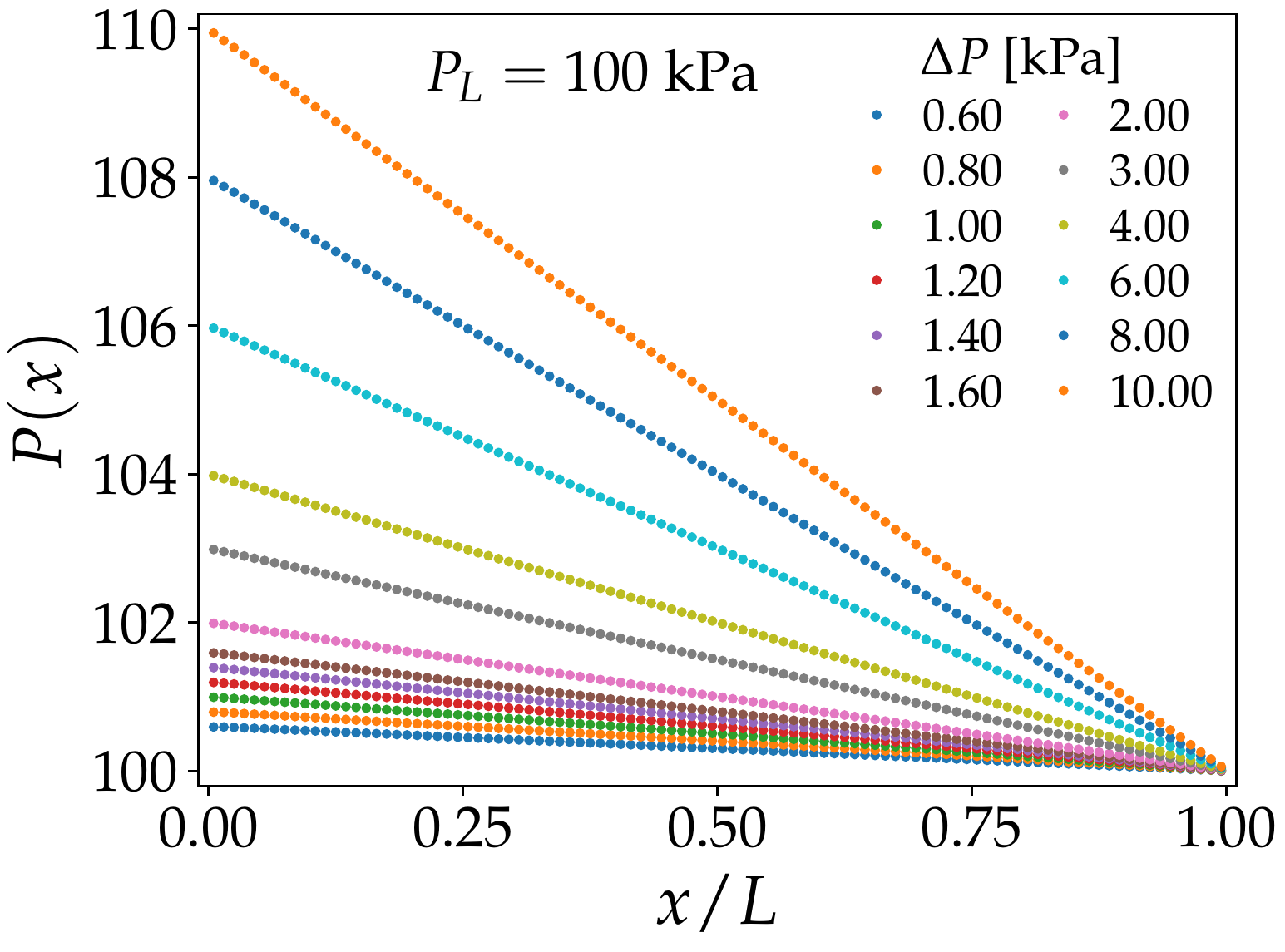}\hfill}
  \centerline{\hfill (a) \hfill \hfill (b) \hfill}
  \caption{\label{fig_pcuscl}Variation of the pressure $P(x)$ [kPa]
    inside a compressible bubble along the tube during steady state
    flow. $P(x)$ shows a linear behavior for different values of
    $\Delta P$ and $P_L$.}
\end{figure}

Figure \ref{fig_gcuscl} shows the variation of $G_{\rm C}(x)$ along
the tube for two different outlet pressures, $P_L=1\,{\rm kPa}$ and
$100\,{\rm kPa}$ where we plot the results for the same set of
pressure drops $\Delta P$. These results are with zero surface
tension, $\gamma=0$. There are a few details to note here. First, the
plots show that $G_{\rm C}(x)$ increases with an increase in $\Delta
P$. In addition, $G_{\rm C}(x)$ also depends on the absolute pressures
at the inlet and outlet, since we can see that the curves are
non-linear functions of $x$ for $P_L=1\,{\rm kPa}$, whereas for
$P_L=100\,{\rm kPa}$, they show linear behavior. Furthermore, $G_{\rm
  C}(x)$ approaches $\Delta P/P_L$ at $x=L$ for all the data sets.

To explain the dependency of $G_{\rm C}(x)$ on $\Delta P$ and $P_L$,
we recall Equation \ref{eqn_pcfl} and rewrite Equation \ref{eqn_gcdef}
as,
\begin{equation}
  \label{eqn_gctop}
  \displaystyle
  G_{\rm C}(x) = \frac{P_0-P(x)}{P(x)} \;,
\end{equation}
where $P(x)$ is the pressure inside a bubble at $x$. For $x=L$, $P(x)
= P_L$ and therefore $G_{\rm C}(L)=\Delta P/P_L$ as observed. In
Figure \ref{fig_pcuscl}, we plot $P(x)$, averaged over different time
steps in the steady state, for the two outlet pressures, $P_L=1\,{\rm
  kPa}$ and $100\,{\rm kPa}$. Both of the plots show linear variation
along $x$ with the slope $-\Delta P$. We therefore have $P(x)=-x\Delta
P/L+P_L+\Delta P$ and thus,
\begin{equation}
  \label{eqn_gcscl}
  \displaystyle
  \frac{G_{\rm C}(x)}{n_P} = \frac{x/L}{1+n_P(1-x/L)} \;,
\end{equation}
where $n_P=\Delta P/P_L$. This leads to
\begin{equation}
  \label{eqn_gclim}
  \displaystyle
  \frac{G_{\rm C}(x)}{n_P} \sim
  \begin{cases}
    \displaystyle
    \frac{1}{n_P}\left(\frac{1}{1-x/L}-1\right) & {\rm for} \quad n_P \gg 1 \;, \\
    x/L                                         & {\rm for} \quad n_P \ll 1 \;,
  \end{cases}
\end{equation}
which explains the concave and linear variation of $G_C$ as function
of $x/L$ observed in Figure \ref{fig_gcuscl} (a) and (b)
respectively. The growth of the bubbles along the tube is therefore a
function of $n_P=\Delta P/P_L$.

\begin{figure}[htbp]
  \centerline{\hfill
    \includegraphics[width=0.45\textwidth]{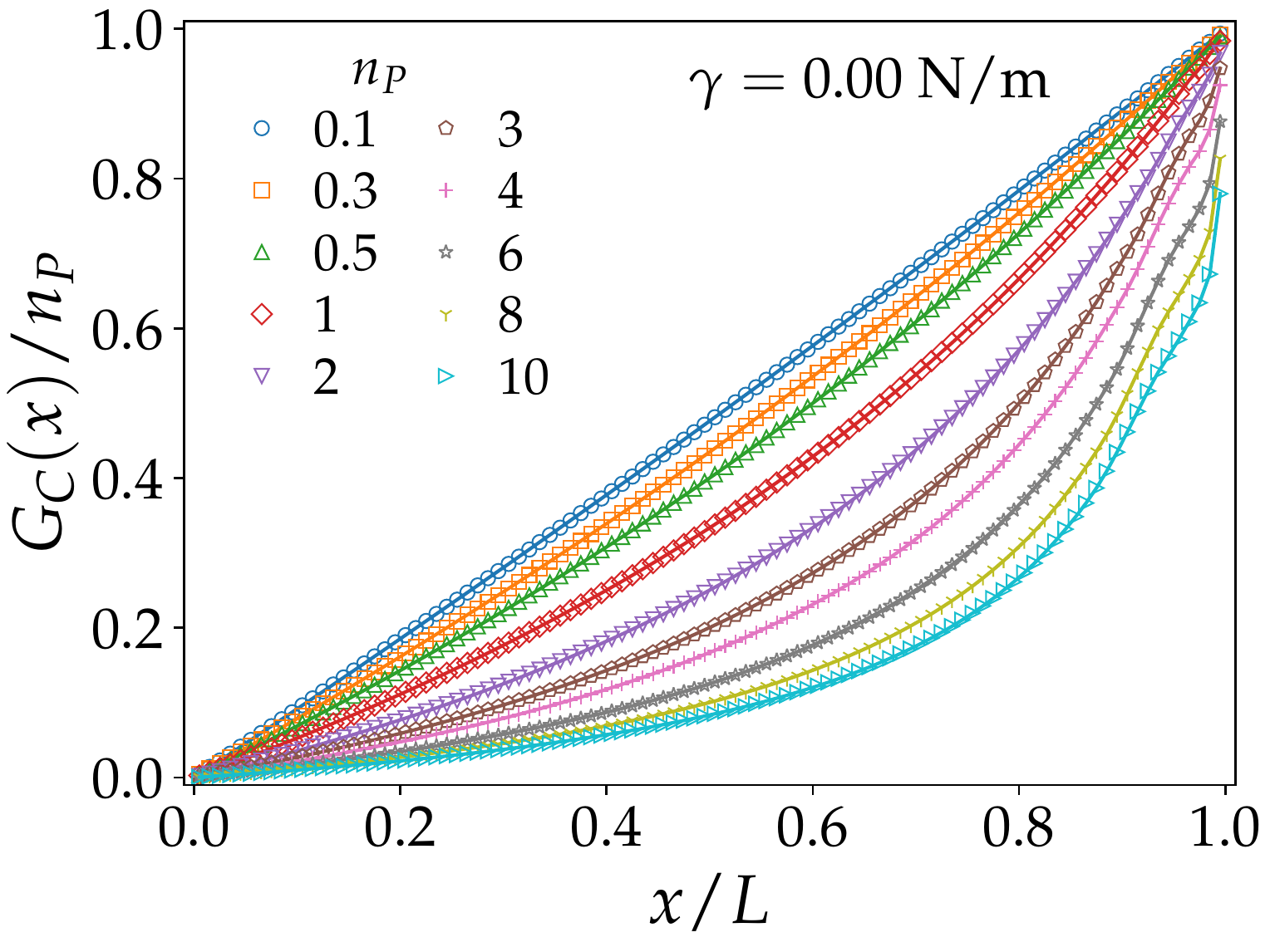}\hfill
    \includegraphics[width=0.45\textwidth]{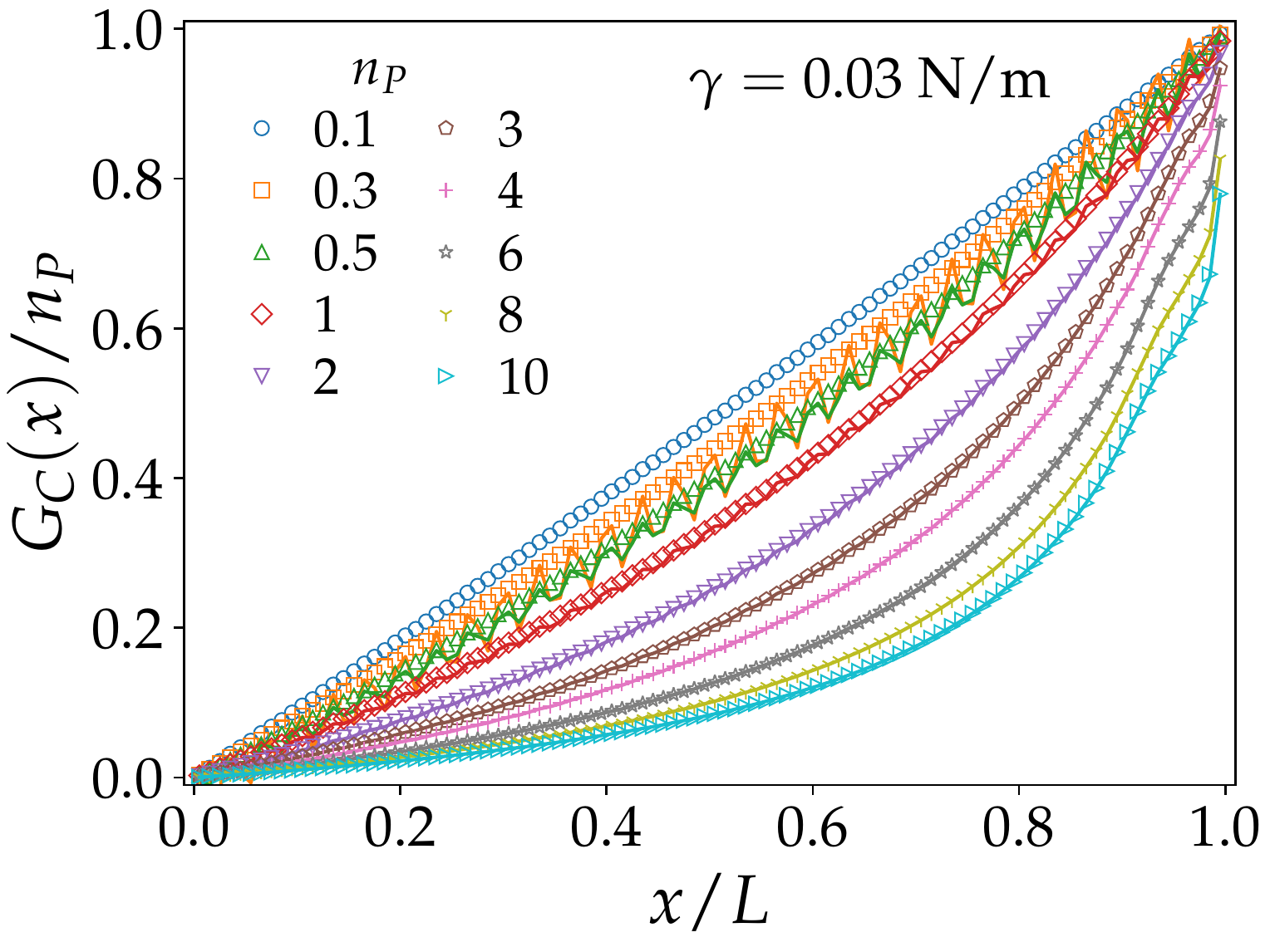}\hfill}
  \centerline{\hfill (a) \hfill \hfill (b) \hfill}
  \caption{\label{fig_gcscl}Variation of the bubble growth $G_{\rm
      C}$, scaled with $n_P=\Delta P/P_L$, with $x/L$. Results are
    plotted for the same sets of $n_P$ for two different values of
    $P_L$. The left and right figures correspond to $\gamma=0$ and
    $0.03\,{\rm N/m}$ respectively. In each plot, the line corresponds
    to $P_L=1\,{\rm kPa}$ and the symbols correspond to $P_L=100\,{\rm
      kPa}$.}
\end{figure}

In Figure \ref{fig_gcscl} we plot $G_{\rm C}/n_P$ for the two outlet
pressures $P_L$ with the same sets of values of $n_P$ for (a) $\gamma
= 0$ and (b) $\gamma = 0.3\,{\rm N/m}$. The plots show that the
results for the same values of $n_P$ follow the same curves,
irrespective of the outlet pressures $P_L$. Furthermore, for the
non-zero surface tension case in Figure \ref{fig_gcscl} (b), $G_{\rm
  C}$ also shows a periodic oscillation along $x$ when both the $n_P$
and $P_L$ are small, that is, for $P_L=1\,{\rm kPa}$ and $n_P\le
1$. In addition, there is no data point for $n_P\le 0.3$ with
$P_L=1\,{\rm kPa}$, as the movement of the bubbles stopped due to high
capillary barriers. This suggests the existence of an effective
threshold pressure, below which there will be no flow through the
tube. This threshold depends on both $\gamma$ and $P_L$, which we will
explore more in the following section. We show the different
characteristics of flow in the videos provided in electronic
supplementary material.

\subsection{Effective rheology}
\label{effective}

Equations \ref{eqn_pcfl} and \ref{eqn_qifl} resist analytical
solutions even in the case when there is only a single compressible
bubble in the tube. This is due to the pressure in the compressible
bubble being inversely proportional to the difference in position of
the two menisci surrounding it, whereas the motion of the two
surrounding incompressible fluids is determined by the cosine of the
positions of the same menisci. These equations, even in this simplest
case, are therefore highly non-linear with an essential singularity
lurking in the very neighborhood where we seek solutions. We therefore
stick to numerical analysis in the following.

\begin{figure}[htbp]
  \centerline{\hfill
    \includegraphics[width=0.45\textwidth]{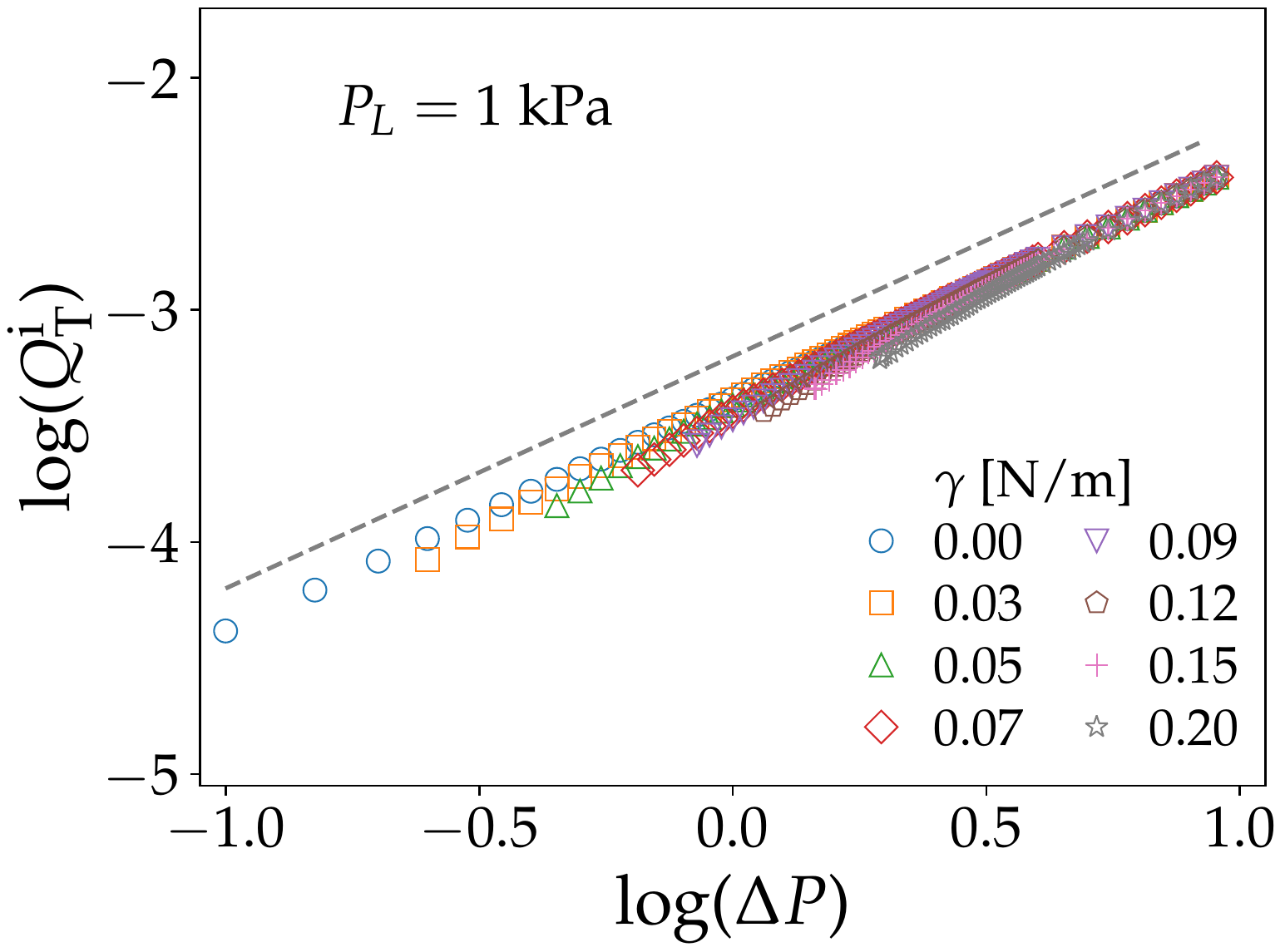}\hfill
    \includegraphics[width=0.45\textwidth]{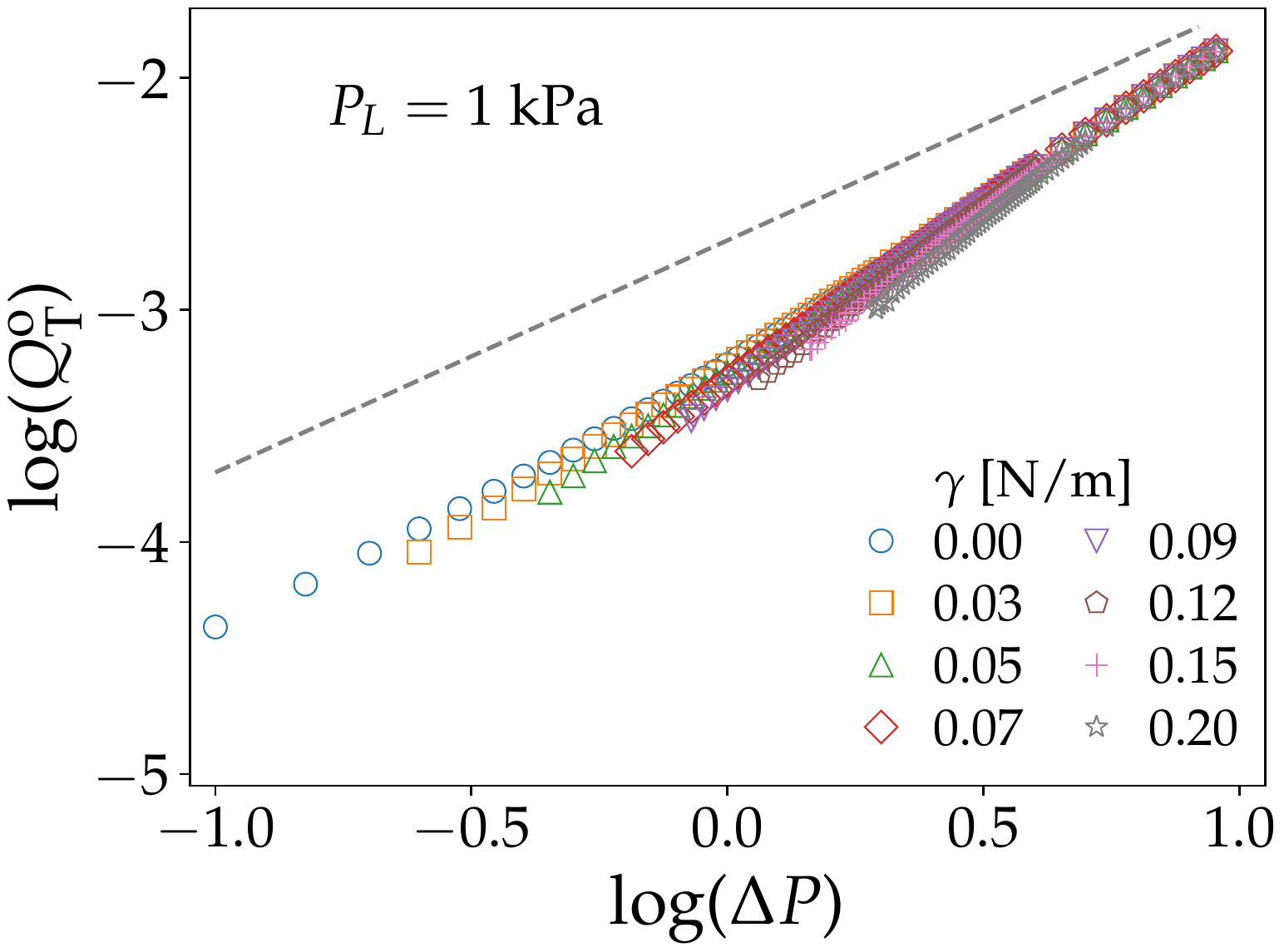}\hfill}
  \centerline{\hfill
    \includegraphics[width=0.45\textwidth]{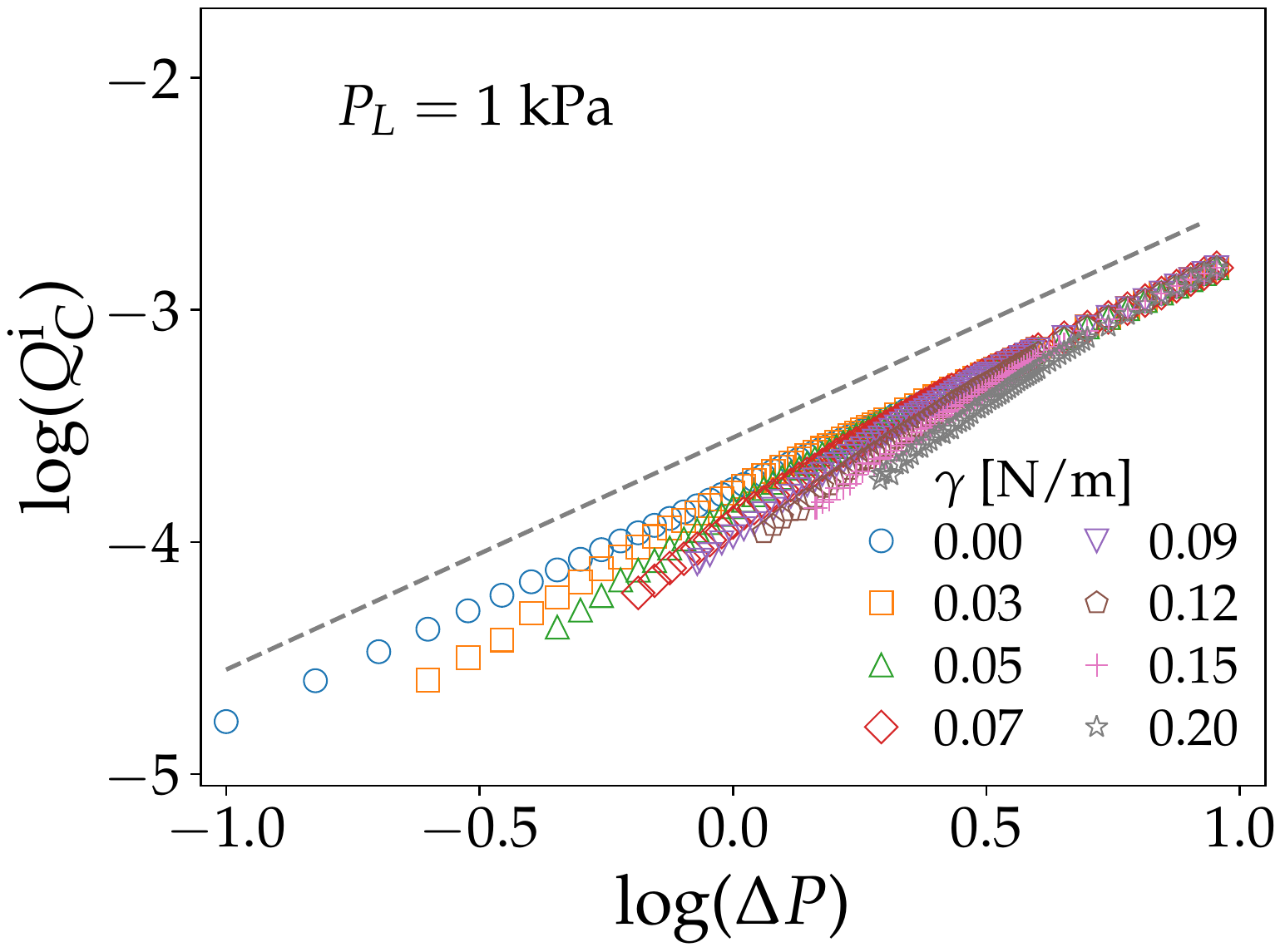}\hfill
    \includegraphics[width=0.45\textwidth]{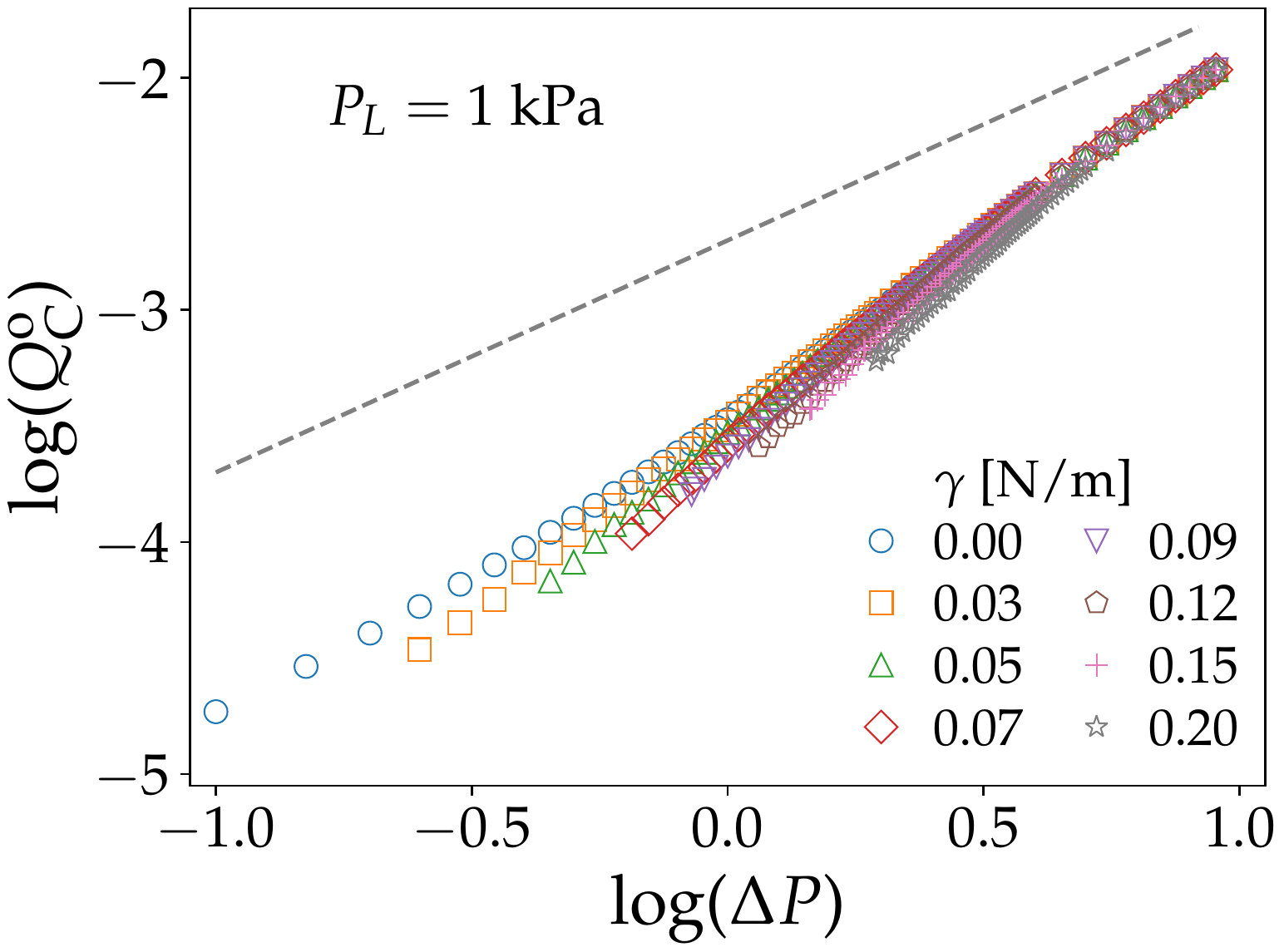}\hfill}
   \centerline{\hfill
    \includegraphics[width=0.45\textwidth]{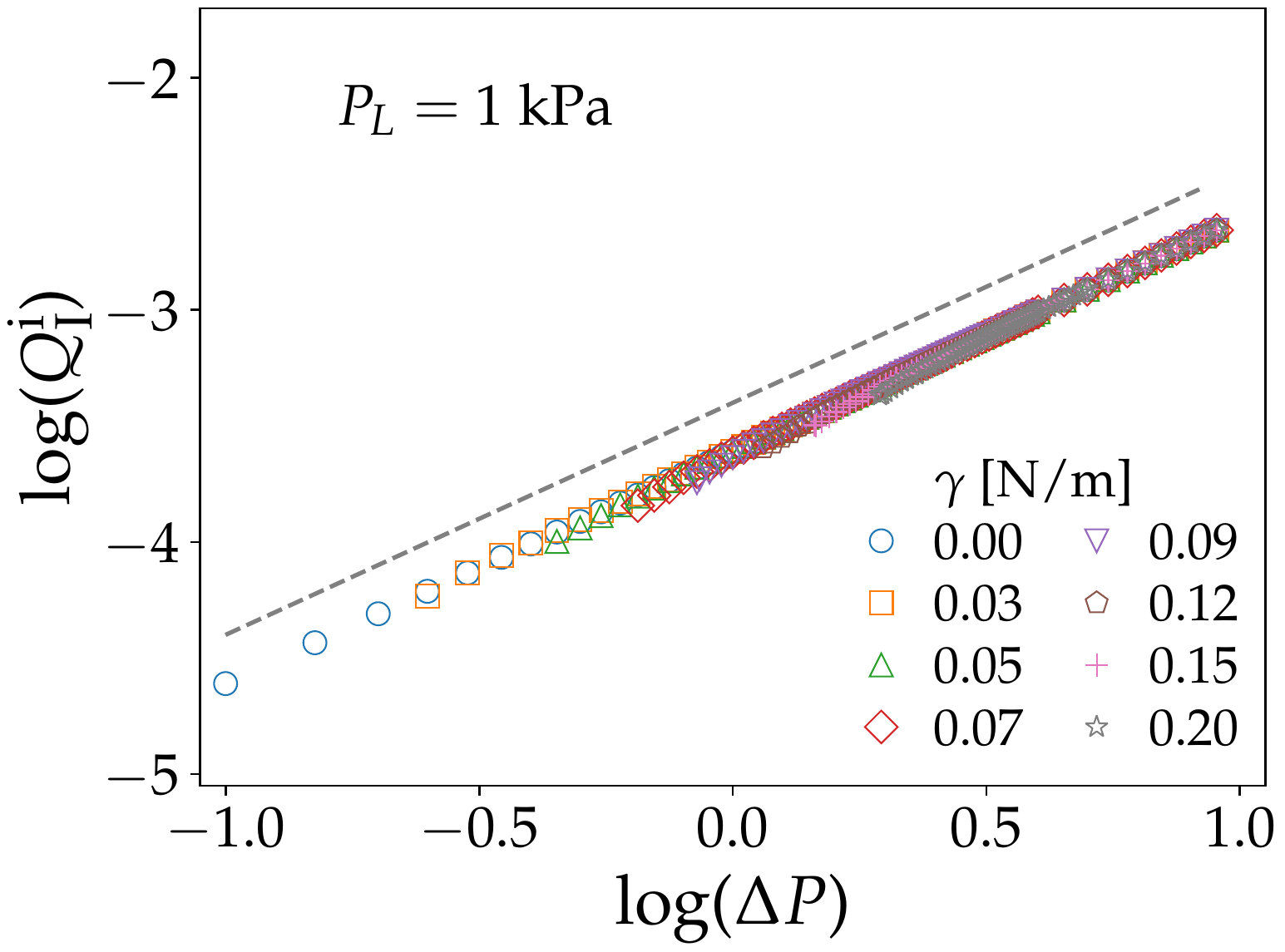}\hfill
    \includegraphics[width=0.45\textwidth]{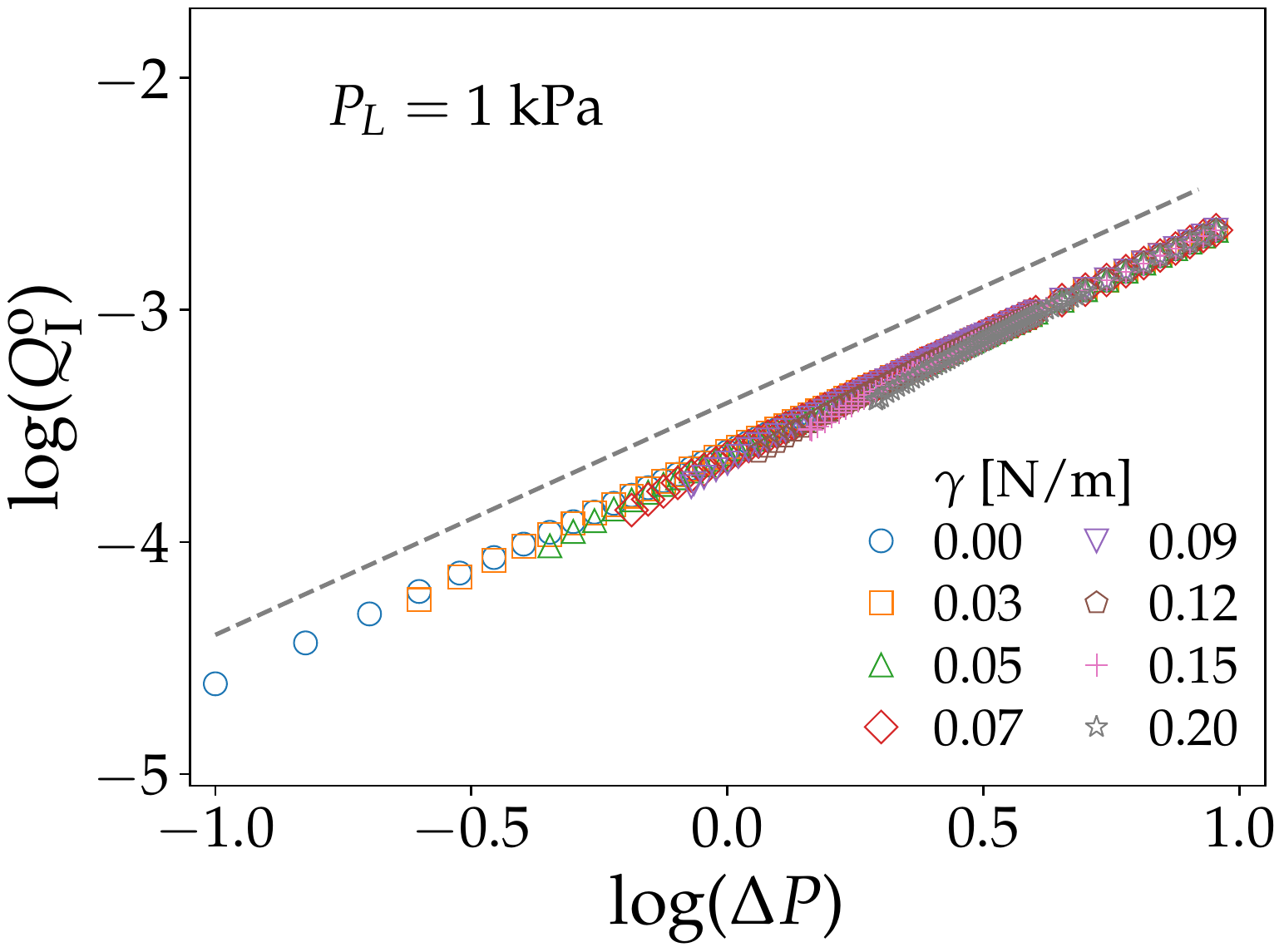}\hfill}
  \caption{\label{fig_qinout} Plot of the flow rates for the total
    ($Q_{\rm T}^{\rm i,o}$), compressible ($Q_{\rm C}^{\rm i,o}$) and
    incompressible ($Q_{\rm I}^{\rm i,o}$) fluids at the inlet (left
    column) and at the outlet (right column) for $P_L=1\,{\rm kPa}$ as
    a function of $\Delta P$. The different sets in each plot
    correspond to different values of the surface tension indicated in
    the legends. The dashed line in each plot has a slope of $1$.}
\end{figure}

Due to the volumetric growth of the compressible bubbles during their
flow towards the outlet, the volumetric flow rate varies along the
tube. In addition, this volumetric growth is a function of the
pressures, making the average saturation and the effective viscosity
of the two fluids inside the tube pressure dependent. These two
mechanisms together control the effective rheological behavior of the
steady-state flow. In Figure \ref{fig_qinout} we show the variation of
the volumetric flow rates ($Q_{\rm T}^{\rm i,o}$, $Q_{\rm C}^{\rm
  i,o}$, $Q_{\rm I}^{\rm i,o}$) as functions of the pressure drop
$\Delta P$ for the outlet pressure $P_L=1\,{\rm kPa}$ and for
different values of the surface tension ($\gamma$). Note the
differences between the inlet and outlet flow rates for the total and
for the each component of flow. For the incompressible fluid, there is
no significant increase in the outlet flow rate compared to its inlet
flow rate (third row in Figure \ref{fig_qinout}) whereas there is a
noticeable increase in the outlet flow rate of the compressible fluid
(second row in Figure \ref{fig_qinout}). This increase in $Q_{\rm
  C}^{\rm o}$ effectively increases the total flow rate at the outlet
(first row in Figure \ref{fig_qinout}). The dashed line in Figure
\ref{fig_qinout} has a slope equal to $1$. The total flow rates show
deviations from this dashed line. For the inlet, $Q_{\rm T}^{i}$ shows
small deviations from the dashed line for $\gamma>0$ at small $\Delta
P$. Whereas at the outlet, the deviations are significantly higher due
to the increase in the volumetric growth of the compressible fluid.

Another point to note in Figure \ref{fig_qinout} is that there is a
minimum value of $\Delta P$, below which there is no data point
available. This is due to the existence of a threshold pressure below
which the flow stops. In the supplementary material we show a
simulation video in this regime where one can observe that the flow of
the bubbles stops at a certain time step. The threshold is due to the
capillary forces at the menisci between the two fluids that create
capillary barriers at the narrowest points along the tube. Such
threshold was also observed in the case of two-phase flow of two
incompressible fluids in a tube with variable radius
\cite{shb13}. There, it was shown analytically that the average flow
rate $Q$ in the steady state varies with the applied pressure drop
$\Delta P$ as, $Q\sim\sqrt{\Delta P^2 - P_{\rm t}^2}$ where $P_{\rm
  t}$ is the effective threshold pressure. When $\abs{\Delta P} -
P_{\rm t} \ll P_{\rm t}$, this relationship translates to
$Q\sim\sqrt{\abs{\Delta P} - P_{\rm t}}$, that is, the flow rate
varies with the excess pressure drop to the power of $0.5$. The
threshold pressure depends on the surface tension and on the
configuration of the menisci positions inside the tube. If the total
capillary barrier is higher than the applied pressure drop, the flow
stops. This is similar here for the two-phase flow with one of the
fluids being compressible.

\begin{figure}[htbp]
  \centerline{\hfill
    \includegraphics[width=0.33\textwidth]{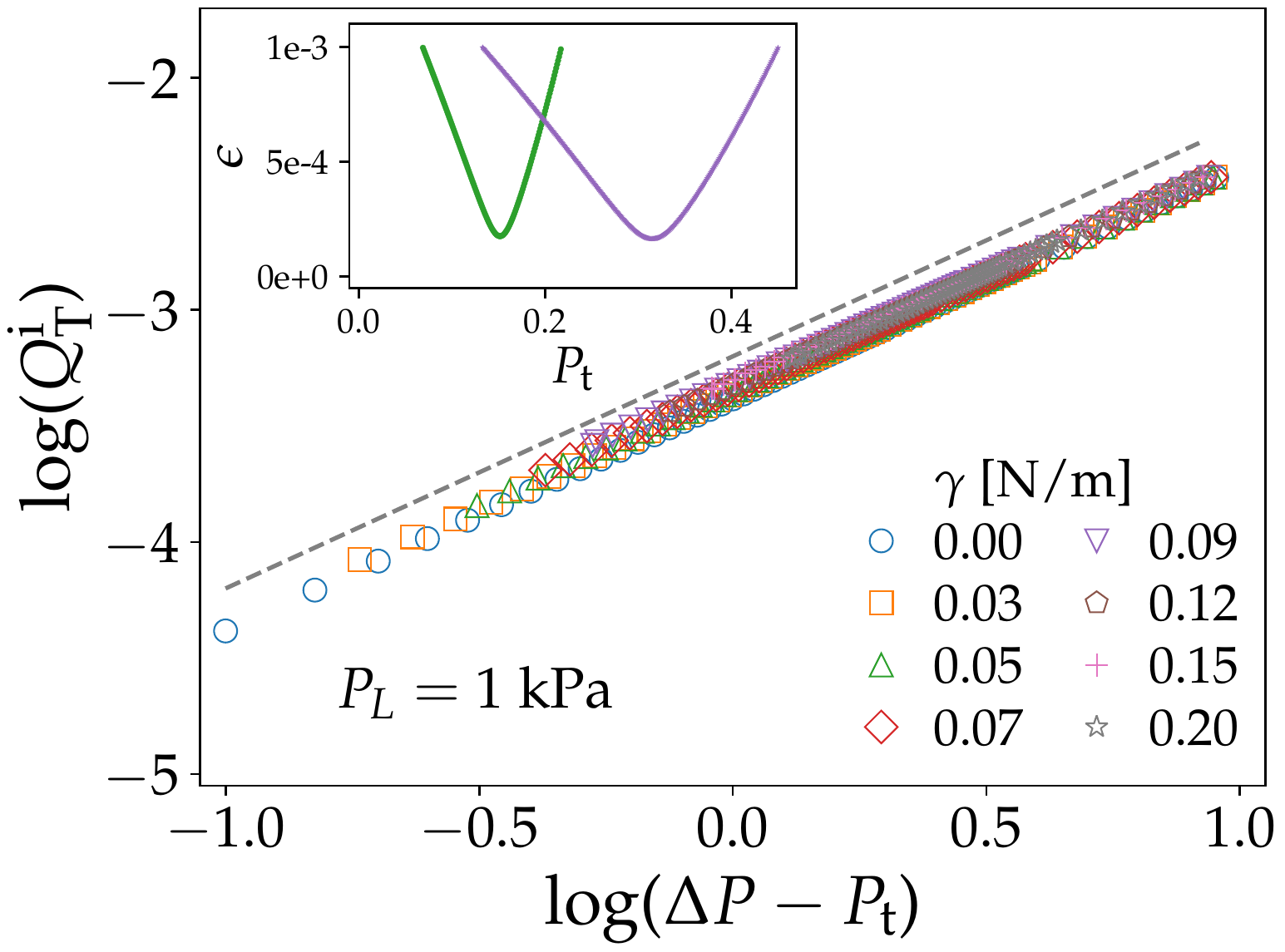}\hfill
    \includegraphics[width=0.33\textwidth]{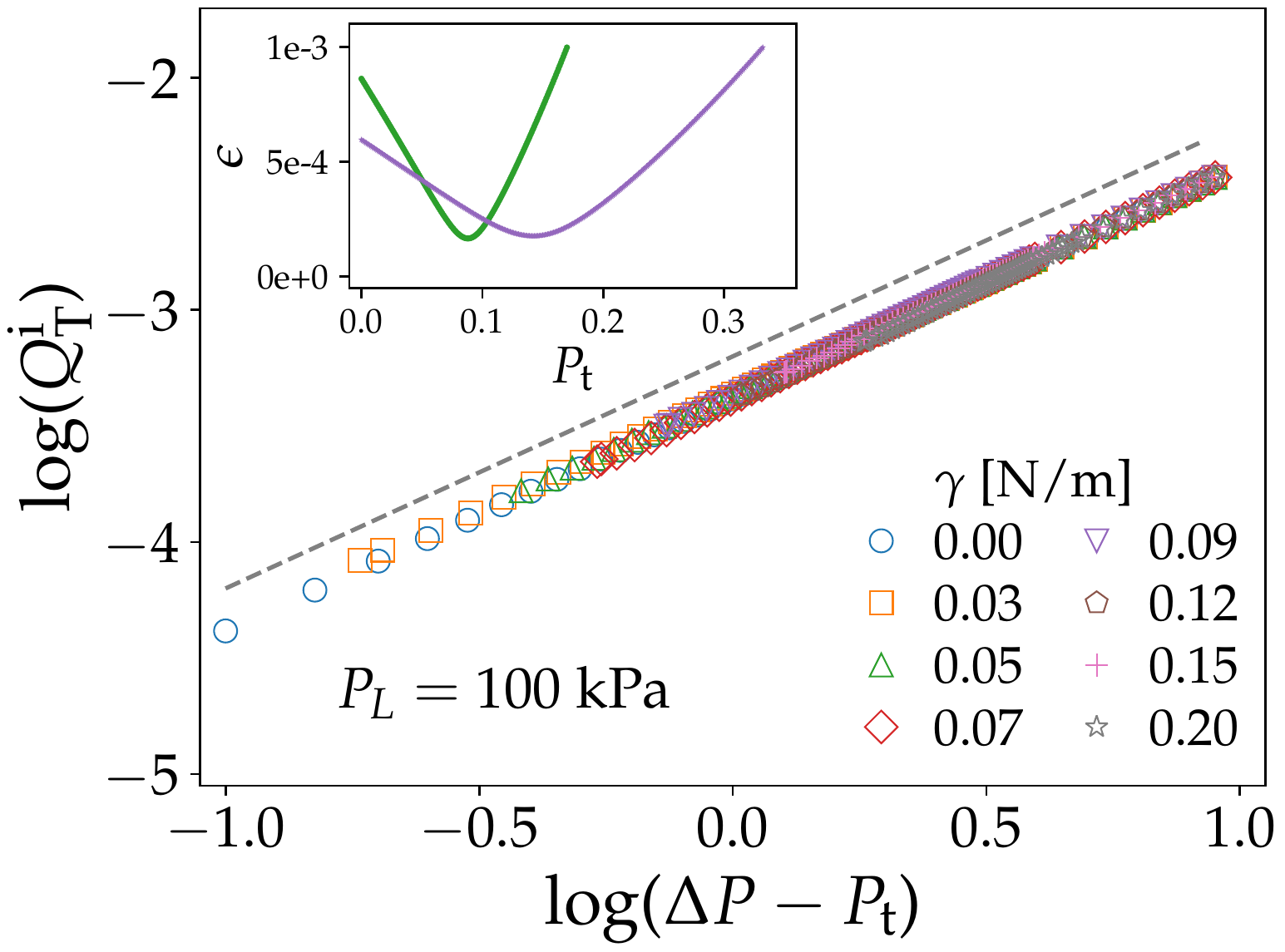}\hfill
    \includegraphics[width=0.33\textwidth]{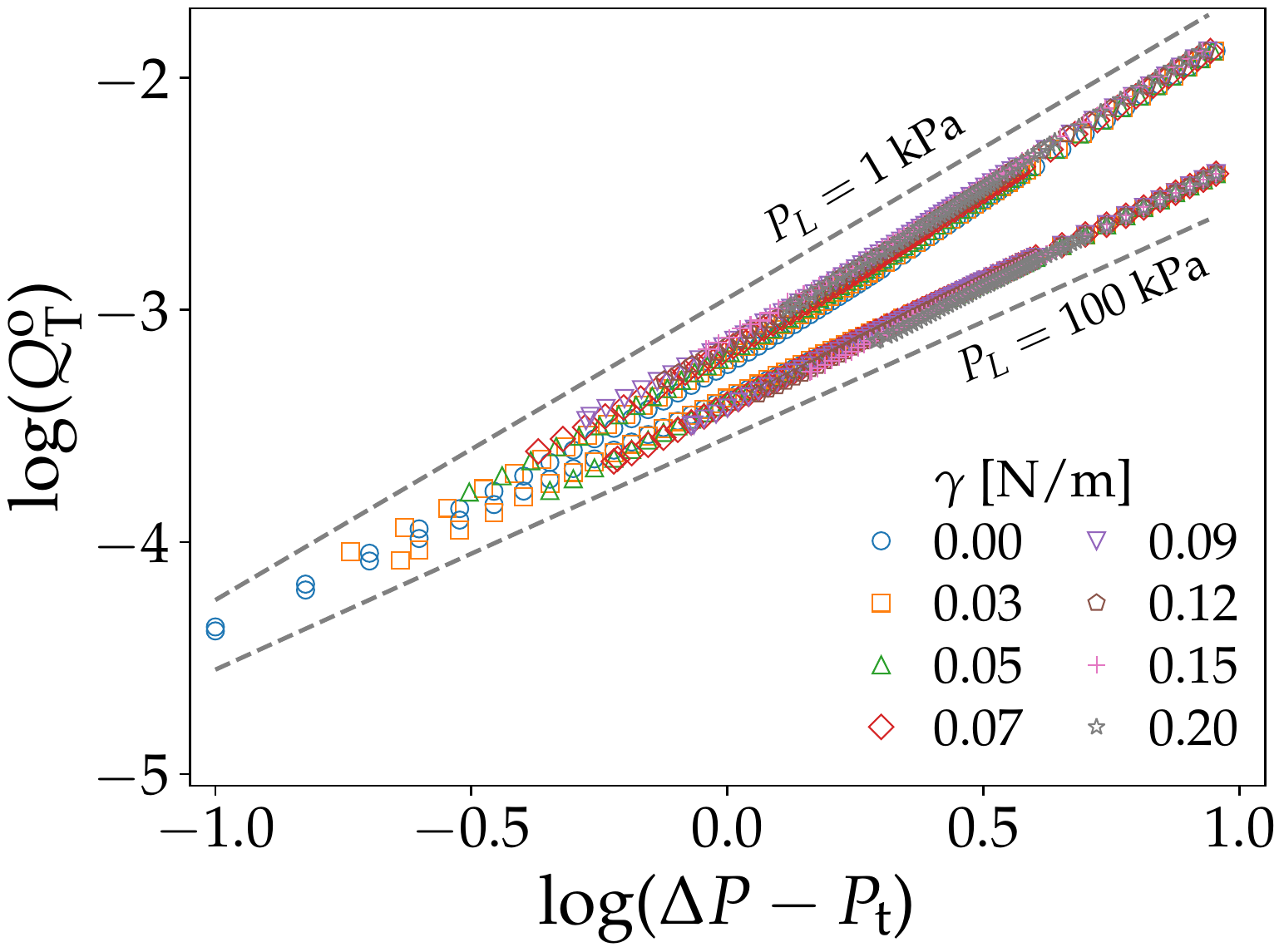}\hfill}
  \centerline{\hfill (a) \hfill \hfill (b) \hfill \hfill (c) \hfill}
  \caption{\label{fig_qfit}Plot of the volumetric inlet flow rate
    $Q_{\rm T}^{\rm i}$ as a function of the excess pressure drop
    $(\Delta P-P_{\rm t})$ for $P_L=1\,{\rm kPa}$ and $100\,{\rm
      kPa}$, where the values of $P_{\rm t}$ are obtained from a
    minimization of the least square fit error $\epsilon$. The
    minimization is illustrated in the insets of (a) and (b) for
    $\gamma=0.05\,{\rm N/m}$ (green) and $0.09\,{\rm N/m}$
    (purple). The dashed lines in (a) and (b) have a slope $1$ whereas
    in (c), the lower and upper dashed lines have slopes $1$ and $1.3$
    respectively.}
\end{figure}

We assume a general relation between the average volumetric flow rates
$Q_{\rm T}^{\rm i,o}$ and the pressure drop $\Delta P$ as,
\begin{equation}
  \label{eqn_pqbeta}
  \displaystyle Q_{\rm T}^{\rm i,o} \sim (\Delta P - P_{\rm t})^{\beta_{\rm i,o}}
\end{equation}
where $\beta_{\rm i,o}$ is the corresponding exponent. In order to
find both the effective threshold pressure $P_T$ and the exponent
$\beta_{\rm i,o}$ from the measurements of $Q_{\rm T}^{\rm i,o}$, we
adopt an error minimization technique that was used in earlier studies
\cite{sh12,fsr21}. There we choose a series of trial values for
$P_{\rm t}$ and calculate the mean square error $\epsilon$ for the
linear least square fit by fitting the data points with
$\log(Q)\sim\log(\Delta P-P_{\rm t})$. Then we select the value of
$P_{\rm t}$ that corresponds to the minimum value of $\epsilon$,
implying the best fit of the data points with Equation
\ref{eqn_pqbeta}. This is illustrated in the insets of Figure
\ref{fig_qfit} (a) and (b). The slope for the selected threshold
$P_{\rm t}$ provides the exponent $\beta_{\rm i,o}$. The variation of
the total inlet and outlet flow rates $Q_{\rm T}^{\rm i,o}$ with the
excess pressure drop $(\Delta P-P_{\rm t})$ are plotted in Figure
\ref{fig_qfit} for the two outlet pressures $P_L=1$ and $100\,{\rm
  kPa}$. The data sets show agreement with Equation \ref{eqn_pqbeta}
with the selected values of $P_{\rm t}$ and $\beta$. There is a
noticeable difference between the slopes for the inlet and outlet flow
rates for $P_L=1\,{\rm kPa}$ whereas for $P_L=100\,{\rm kPa}$ they are
similar. For $P_L=100\,{\rm kPa}$ the data points for both $Q_{\rm
  T}^{\rm i}$ and $Q_{\rm T}^{\rm o}$ follow a slope of $\approx 1.0$
whereas for $P_L=1\,{\rm kPa}$, the data points for $Q_{\rm T}^{\rm
  i}$ and $Q_{\rm T}^{\rm o}$ follow the slopes of $\approx 1.0$ and
$1.3$ respectively. These are indicated by the dashed lines in the
figures.

\begin{figure}[htbp]
  \centerline{\hfill
    \includegraphics[width=0.33\textwidth]{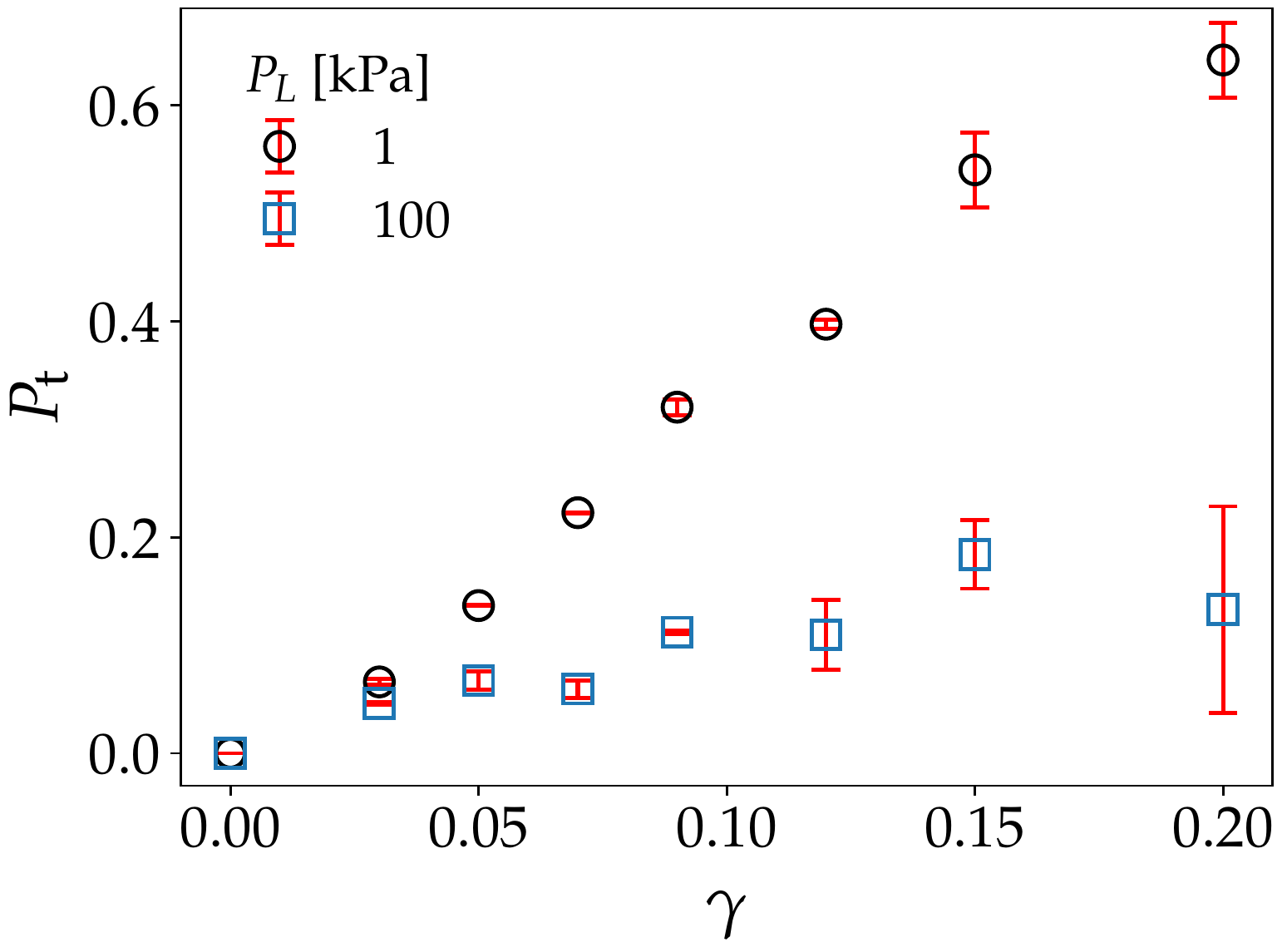}\hfill
    \includegraphics[width=0.33\textwidth]{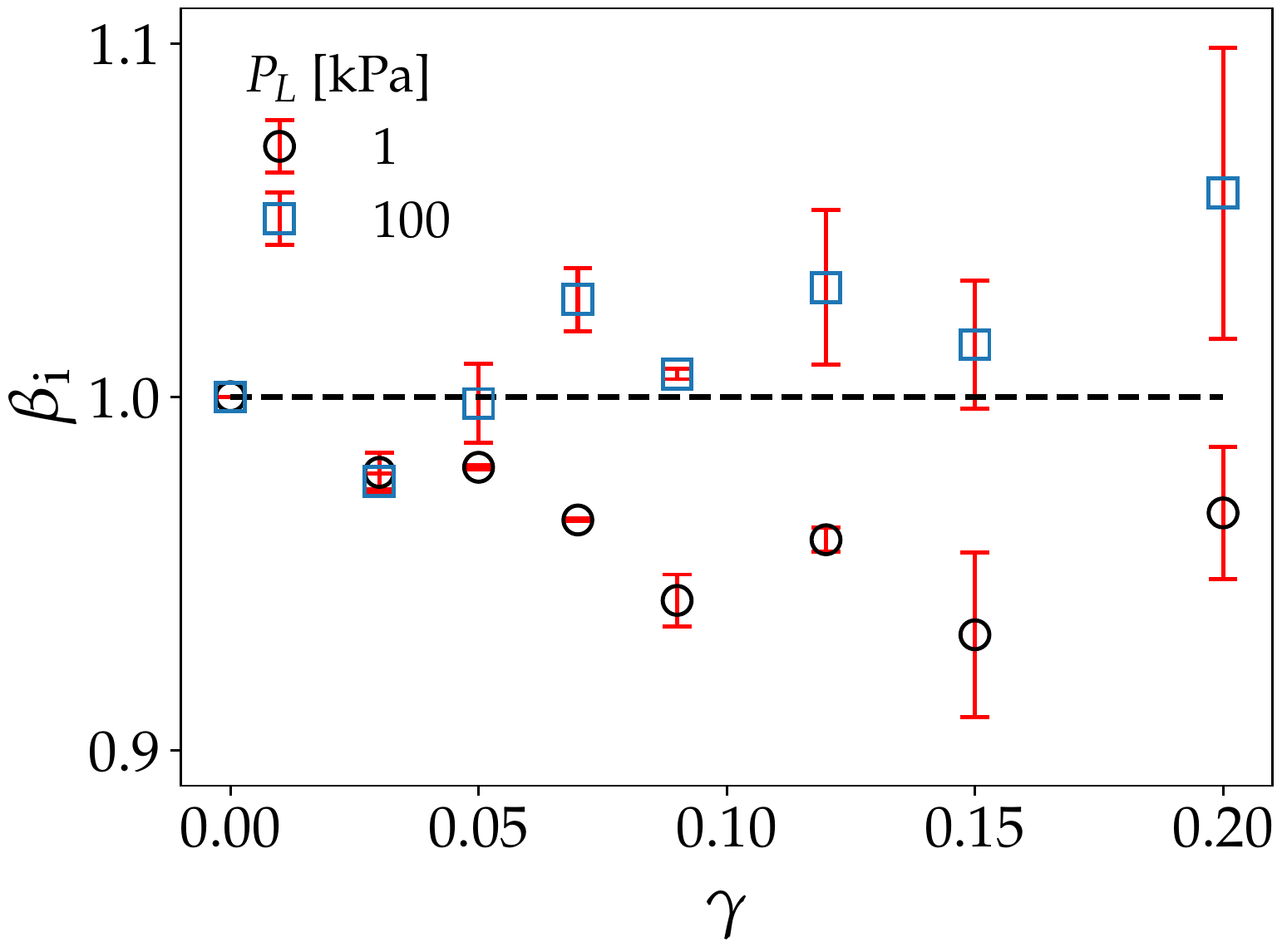}\hfill
    \includegraphics[width=0.33\textwidth]{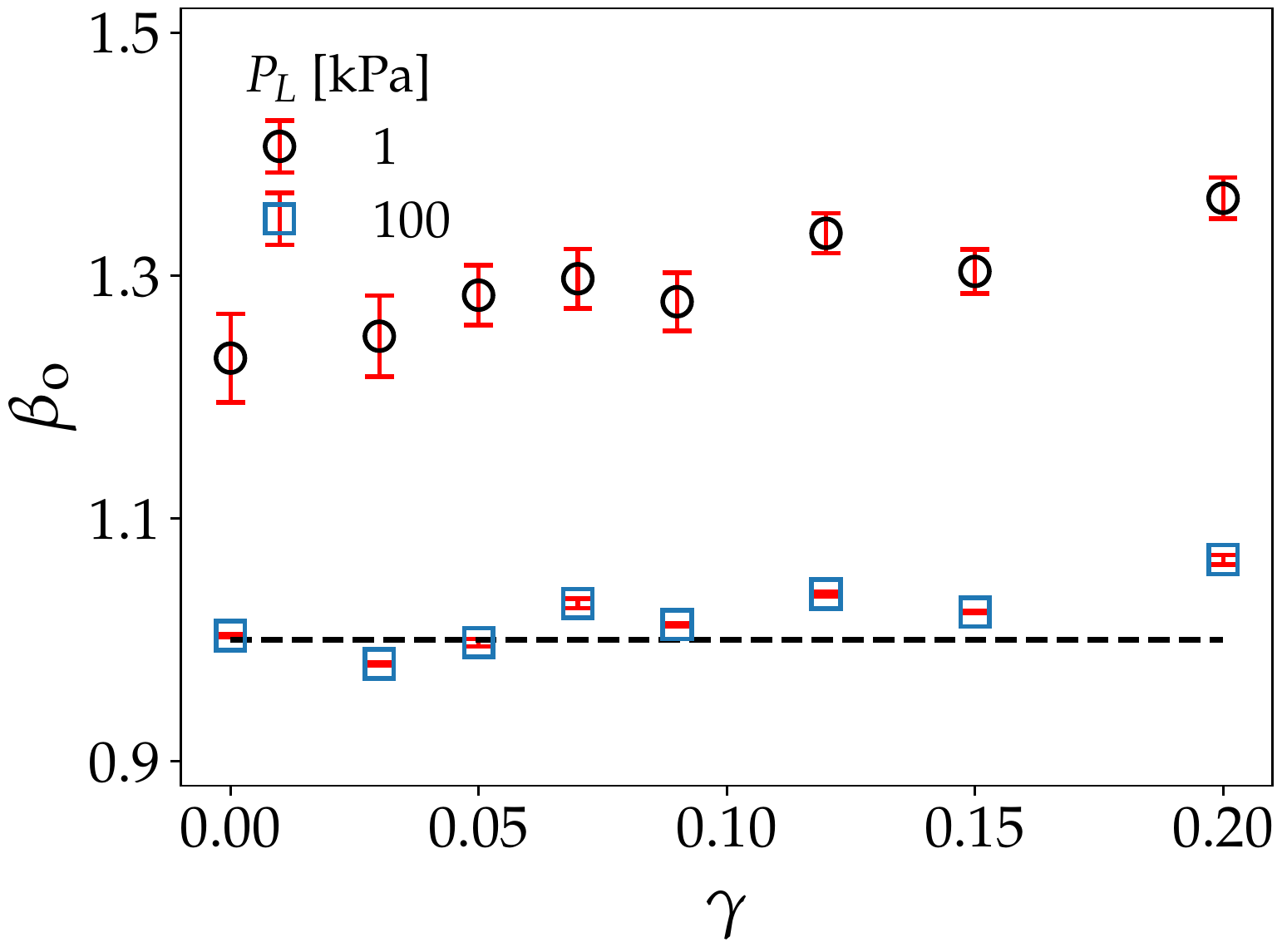}\hfill}
  \centerline{\hfill (a) \hfill \hfill (b) \hfill \hfill (c) \hfill}
  \caption{\label{fig_betapt}Variation of the threshold pressure
    $P_{\rm t}$ and the exponents $\beta_{\rm i,o}$ as functions of
    the effective surface tension $\gamma$ for $P_L=1$ and $100\,{\rm
      kPa}$. $P_{\rm t}$ increases with the increase of $\gamma$ and
    the values are much higher for $P_L=1\,{\rm kPa}$ compared to
    $P_L=100\,{\rm kPa}$. The exponent $\beta_{\rm i}$ for the inlet
    flow rate are close to $1$ for both the values of $P_L$ whereas
    for the outlet flow rate $\beta_{\rm o}\approx 1.3$ for
    $P_L=1\,{\rm kPa}$. For $P_L=100\,{\rm kPa}$, $\beta_{\rm o}$
    remains close to $\beta_{\rm i}$. The dashed horizontal lines
    indicate the value $1.0$ of the $y$ axis.}
\end{figure}

The variations of $P_{\rm t}$ and $\beta_{\rm i,o}$ with the surface
tension $\gamma$ are plotted in Figure \ref{fig_betapt}. The data
points were calculated by considering different ranges of $\Delta P$
and taking averages over the ranges, and the corresponding standard
deviations are plotted as error bars. The threshold pressure $P_{\rm
  t}$ is zero at $\gamma=0$ and then increases gradually with $\gamma$
which shows that the threshold appears due to capillary forces. The
increase in $P_{\rm t}$ with $\gamma$ appears to be linear here which
is similar to the case of two incompressible fluids, where the linear
dependence of $P_{\rm t}$ on the surface tension was shown
analytically \cite{shb13}. Additionally for the compressible flow
here, the thresholds also depend on the outlet pressure $P_L$. For the
lower outlet pressure $P_L=1\,{\rm kPa}$, the thresholds are
systematically higher compared to those for $P_L=100\,{\rm kPa}$ for
the whole range of $\gamma$. Furthermore, the exponents $\beta_{\rm
  i,o}$ also depend on the outlet pressure as seen from the Figures
\ref{fig_betapt} (b) and (c). The difference is more visible for the
exponents related to the outlet flow rates than the inlet. For the
inlet flow rate, $\beta_{\rm i}$ has values around $\approx 0.95$ and
$1.02$ for $P_L=1\,{\rm kPa}$ and $100\,{\rm kPa}$ respectively,
showing almost linear dependence for both the cases. For the outlet
flow rates, $\beta_{\rm o}$ remains close to $\beta_{\rm i}$ for
$P_L=100\,{\rm kPa}$ whereas for $P_L=1\,{\rm kPa}$, $\beta_{\rm o}$
increases to $\approx 1.3$. This increase in $\beta_{\rm o}$ compared
to $\beta_{\rm i}$ reflects the dependence of the volumetric growth
$G_{\rm C}(x)$ of the bubbles on $P_L$, indicating an underlying
dependence of the rheological behavior on the absolute inlet or outlet
pressures. However, at this point we are unable to describe how the
two parameters $P_{\rm t}$ and $\beta$ scale with $P_L$, which needs
further study.

Existing studies of the power-law volumetric flow rate-pressure drop
relation for porous media have shown the existence of different
regimes characterized by different exponents.  These studies involve
experiments \cite{tkr09, tlk09, rcs11, sbd17, glb20, zbg21}, Lattice
Boltzmann simulations \cite{yts13}, pore-network modeling \cite{sh12,
  sbd17} and analytical calculations \cite{tlk09, sh12, rhs19,
  zbg21}. There are three regimes, an intermediate non-linear regime
where the flow rate $Q$ increases at a rate much faster than the
applied pressure drop $\Delta P$ with a power law exponent larger than
one and up to a value around 2.5. There are in addition two linear
regimes for either smaller \cite{yts13, glb20, zbg21} or larger
\cite{yts13, sh12,sbd17} volumetric flow rates than the non-linear
regime. This allows the definition of a lower and upper crossover
pressure drop. A simple explanation for these three regimes may be
drawn from the study of the conductivity of a disordered network of
threshold resistors \cite{rh87} which is based on the following
idea. Each resistor has a threshold voltage to start conducting the
current and then the current increases linearly. In a network with
links of such flow properties, there will be a regime when new
conducting paths will appear when increasing the global pressure
drop. The increase in the flow rate through each path together with
the increase in the number of paths leads to an effective increase of
$Q$ faster than $\Delta P$. This results in the non-linear exponent
being higher than $1$, the value of which depends on the distribution
of the thresholds in each link \cite{rhs19}. The linear regime above
this non-linear regime appears from all the available paths being
conducting whereas the linear regime below appears from the flow being
flow in single percolating channels without interfaces. With this
idea, the experimental \cite{glb20, zbg21} and numerical \cite{yts13}
observations of two-phase flow in porous media showing linear
variation of flow rate in this low pressure regime therefore indicate
that the flow in single channels consisting of many pores are linear,
which is similar to what we have found for the lower outlet pressure
in the present compressible/incompressible flow case.

\section{Conclusions}
\label{conclusion}

We have studied the flow of alternating compressible bubbles and
incompressible droplets through a capillary tube with variable
radius. The motion of the bubbles was given by the model Equations
\ref{eqn_pcfl} and \ref{eqn_qifl}, thus assuming the compressible
fluid to be an ideal gas with zero viscosity, whereas the
incompressible fluid is Newtonian. The incompressible fluid is more
wetting than the compressible gas, but not to a degree that films
form.

We switch between injecting the compressible and incompressible fluid
at intervals so that the fractional flow rate is essentially constant
at the inlet. We fix a pressure drop along the tube in addition to an
ambient pressure. This creates steady-state flow conditions in the
tube.

The compressible bubbles expand as they move from the higher pressure
region at the inlet towards the lower pressure at the outlet.  This
expansion accelerates the incompressible fluid, thus making the
volumetric flow rate larger at the outlet than at the inlet.  The
lower the ambient pressure is, the stronger this effect is.

We measure volumetric flow rate at the inlet, finding essentially a
linear relationship between the volumetric flow rate and the pressure
drop. However, there is a threshold pressure that needs to be overcome
in order to have flow through the tube.

At the outlet, we find that the volumetric flow rate is still linear
in the excess pressure drop when the ambient pressure is low.
However, when the ambient pressure is high, the volumetric flow rate
at the outlet becomes proportional to the excess pressure to a power
of around $1.3$.

This behavior is very different from that of two incompressible fluids
moving through a corresponding tube: Here the volumetric flow rate,
being the same at the inlet and the outlet, is proportional to the
square root of the excess pressure.

We expected the flow rate-pressure drop constitutive relations to be
different in this compressible/incompressible case than that of two
incompressible fluids. However, that we should find {\it linearity\/}
was a big surprise. A precise explanation as to why this is so, is
still lacking.

Besides these surprising results, this work makes a first step in
implementing the modeling of compressible/incompressible fluid
mixtures in dynamic network models. We may then envision using more
sophisticated equations of state for the compressible fluid beyond the
ideal gas law. This allows the consideration of e.g.\ phase
transitions such as boiling and condensation in porous media.

\section{Supplementary material}
\label{suppl}
The electronic supplementary material contains videos showing
different flow characteristics. The videos can be found in the list of
{\it ancillary files} in the arXiv abstract page of this article. We
considered a tube with $L=10\,{\rm cm}$, $w=1\,{\rm cm}$,
$a=0.25\,{\rm cm}$ and $h=5$ (Equation \ref{eqn_tube}). The
simulations were performed for $P_L=1\,{\rm kPa}$ and
$\gamma=0.2\,{\rm N/m}$. The compressible bubbles are colored with
magenta whereas the incompressible droplets are colored with
black. The videos are not in real time. We show four different
simulations with different values of $\Delta P$:

\begin{enumerate}[label=(\alph*)]

\item Flow of a single bubble of compressible gas in incompressible
  fluid. Here $\Delta P = 5\,{\rm kPa}$. The video shows the increase
  in the volume of the bubble as it approaches towards the outlet.

\item Injection of multiple compressible bubbles and incompressible
  droplets at a {\it very low} pressure drop, $\Delta P = 0.3\,{\rm
    kPa}$. The flow stops after a certain time when several interfaces
  appeared in the tube. This shows the existence of a total capillary
  barrier, which is higher than the applied pressure drop here.

\item Two-phase flow of multiple compressible bubbles and
  incompressible droplets at a {\it low} pressure drop, $\Delta P =
  0.4\,{\rm kPa}$. Here the bubbles speed up and slow down as they
  flow, showing the combined effect of the surface tension and the
  shape of the tube. The bubbles also grow in volume towards the
  outlet.

\item Two-phase flow of multiple compressible bubbles and
  incompressible droplets at a {\it higher} pressure drop, $\Delta P =
  3\,{\rm kPa}$. The bubbles do not show any significant slowing down
  in this case, indicating the capillary forces being negligible
  compared to the viscous pressure drop. The volumetric expansion of
  the compressible bubbles can also be observed here.

\end{enumerate}

\section*{Acknowledgments}
We thank Federico Lanza, Marcel Moura and H{\aa}kon Pedersen for
helpful discussions. This work is supported by the Research Council of
Norway through its Centers of Excellence funding scheme, project
number 262644.


\begin{thebibliography}{50}
\ifx \bisbn   \undefined \def \bisbn  #1{ISBN #1}\fi
\ifx \binits  \undefined \def \binits#1{#1}\fi
\ifx \bauthor  \undefined \def \bauthor#1{#1}\fi
\ifx \batitle  \undefined \def \batitle#1{#1}\fi
\ifx \bjtitle  \undefined \def \bjtitle#1{#1}\fi
\ifx \bvolume  \undefined \def \bvolume#1{\textbf{#1}}\fi
\ifx \byear  \undefined \def \byear#1{#1}\fi
\ifx \bissue  \undefined \def \bissue#1{#1}\fi
\ifx \bfpage  \undefined \def \bfpage#1{#1}\fi
\ifx \blpage  \undefined \def \blpage #1{#1}\fi
\ifx \burl  \undefined \def \burl#1{\textsf{#1}}\fi
\ifx \doiurl  \undefined \def \doiurl#1{\url{https://doi.org/#1}}\fi
\ifx \betal  \undefined \def \betal{\textit{et al.}}\fi
\ifx \binstitute  \undefined \def \binstitute#1{#1}\fi
\ifx \binstitutionaled  \undefined \def \binstitutionaled#1{#1}\fi
\ifx \bctitle  \undefined \def \bctitle#1{#1}\fi
\ifx \beditor  \undefined \def \beditor#1{#1}\fi
\ifx \bpublisher  \undefined \def \bpublisher#1{#1}\fi
\ifx \bbtitle  \undefined \def \bbtitle#1{#1}\fi
\ifx \bedition  \undefined \def \bedition#1{#1}\fi
\ifx \bseriesno  \undefined \def \bseriesno#1{#1}\fi
\ifx \blocation  \undefined \def \blocation#1{#1}\fi
\ifx \bsertitle  \undefined \def \bsertitle#1{#1}\fi
\ifx \bsnm \undefined \def \bsnm#1{#1}\fi
\ifx \bsuffix \undefined \def \bsuffix#1{#1}\fi
\ifx \bparticle \undefined \def \bparticle#1{#1}\fi
\ifx \barticle \undefined \def \barticle#1{#1}\fi
\bibcommenthead
\ifx \bconfdate \undefined \def \bconfdate #1{#1}\fi
\ifx \botherref \undefined \def \botherref #1{#1}\fi
\ifx \url \undefined \def \url#1{\textsf{#1}}\fi
\ifx \bchapter \undefined \def \bchapter#1{#1}\fi
\ifx \bbook \undefined \def \bbook#1{#1}\fi
\ifx \bcomment \undefined \def \bcomment#1{#1}\fi
\ifx \oauthor \undefined \def \oauthor#1{#1}\fi
\ifx \citeauthoryear \undefined \def \citeauthoryear#1{#1}\fi
\ifx \endbibitem  \undefined \def \endbibitem {}\fi
\ifx \bconflocation  \undefined \def \bconflocation#1{#1}\fi
\ifx \arxivurl  \undefined \def \arxivurl#1{\textsf{#1}}\fi
\csname PreBibitemsHook\endcsname

\bibitem{b88}
\begin{bbook}
\bauthor{\bsnm{Bear}, \binits{J.}}:
\bbtitle{Dynamics of Fluids in Porous Media}.
\bpublisher{Dover},
\blocation{Mineola, New York}
(\byear{1988})
\end{bbook}
\endbibitem

\bibitem{d92}
\begin{bbook}
\bauthor{\bsnm{Dullien}, \binits{F.A.L.}}:
\bbtitle{Porous Media: Fluid, Transport and Pore Structure}.
\bpublisher{Academic Press},
\blocation{San Diego}
(\byear{1992})
\end{bbook}
\endbibitem

\bibitem{b17}
\begin{bbook}
\bauthor{\bsnm{Blunt}, \binits{M.J.}}:
\bbtitle{Multiphase Flow in Permeable Media}.
\bpublisher{Cambridge University Press},
\blocation{Cambridge}
(\byear{2017})
\end{bbook}
\endbibitem

\bibitem{ffh22}
\begin{bbook}
\bauthor{\bsnm{Feder}, \binits{J.}},
\bauthor{\bsnm{Flekk{\o}y}, \binits{E.G.}},
\bauthor{\bsnm{Hansen}, \binits{A.}}:
\bbtitle{Physics of Flow in Porous Media}.
\bpublisher{Cambridge University Press},
\blocation{Cambridge}
(\byear{2022})
\end{bbook}
\endbibitem

\bibitem{cw85}
\begin{barticle}
\bauthor{\bsnm{Chen}, \binits{J.D.}},
\bauthor{\bsnm{Wilkinson}, \binits{D.}}:
\batitle{Pore-scale viscous fingering in porous media}.
\bjtitle{Phys. Rev. Lett.}
\bvolume{55},
\bfpage{1892}
(\byear{1985}).
\doiurl{10.1103/PhysRevLett.55.1892}
\end{barticle}
\endbibitem

\bibitem{lz85}
\begin{barticle}
\bauthor{\bsnm{Lenormand}, \binits{R.}},
\bauthor{\bsnm{Zarcone}, \binits{C.}}:
\batitle{Invasion percolation in an etched network: Measurement of a fractal
  dimension}.
\bjtitle{Phys. Rev. Lett.}
\bvolume{54},
\bfpage{2226}
(\byear{1985}).
\doiurl{10.1103/PhysRevLett.54.2226}
\end{barticle}
\endbibitem

\bibitem{mfj85}
\begin{barticle}
\bauthor{\bsnm{M{\aa}l{\o}y}, \binits{K.J.}},
\bauthor{\bsnm{Feder}, \binits{J.}},
\bauthor{\bsnm{J{\o}ssang}, \binits{T.}}:
\batitle{Viscous fingering fractals in porous media}.
\bjtitle{Phys. Rev. Lett.}
\bvolume{55},
\bfpage{2688}
(\byear{1985}).
\doiurl{10.1103/PhysRevLett.55.2688}
\end{barticle}
\endbibitem

\bibitem{lmt04}
\begin{barticle}
\bauthor{\bsnm{L{\o}voll}, \binits{G.}},
\bauthor{\bsnm{M\'{e}heust}, \binits{Y.}},
\bauthor{\bsnm{Toussaint}, \binits{R.}},
\bauthor{\bsnm{Schmittbuhl}, \binits{J.}},
\bauthor{\bsnm{M{\aa}l{\o}y}, \binits{K.J.}}:
\batitle{Growth activity during fingering in a porous hele-shaw cell}.
\bjtitle{Phys. Rev. E}
\bvolume{70},
\bfpage{026301}
(\byear{2004}).
\doiurl{10.1103/PhysRevE.70.026301}
\end{barticle}
\endbibitem

\bibitem{zmp19}
\begin{barticle}
\bauthor{\bsnm{Zhao}, \binits{B.}},
\bauthor{\bsnm{MacMinn}, \binits{C.W.}},
\bauthor{\bsnm{Primkulov}, \binits{B.K.}},
\bauthor{\bsnm{Chen}, \binits{Y.}},
\bauthor{\bsnm{Valocchi}, \binits{A.J.}},
\bauthor{\bsnm{Zhao}, \binits{J.}},
\bauthor{\bparticle{et} \bsnm{al}}:
\batitle{Comprehensive comparison of pore-scale models for multiphase flow in
  porous media}.
\bjtitle{Proc. Natl. Acad. Sci. USA.}
\bvolume{116},
\bfpage{13799}
(\byear{2019}).
\doiurl{10.1073/pnas.1901619116}
\end{barticle}
\endbibitem

\bibitem{ltz88}
\begin{barticle}
\bauthor{\bsnm{Lenormand}, \binits{R.}},
\bauthor{\bsnm{Touboul}, \binits{E.}},
\bauthor{\bsnm{Zarcone}}:
\batitle{Numerical models and experiments on immiscible displacements in porous
  media}.
\bjtitle{J. Fluid Mech.}
\bvolume{189},
\bfpage{165}
(\byear{1988}).
\doiurl{10.1017/S0022112088000953}
\end{barticle}
\endbibitem

\bibitem{d56}
\begin{barticle}
\bauthor{\bsnm{Darcy}, \binits{H.}}
\bjtitle{Les Fontaines publiques de la ville de Dijon}
\bvolume{647},
(\byear{1856})
\end{barticle}
\endbibitem

\bibitem{tkr09}
\begin{barticle}
\bauthor{\bsnm{Tallakstad}, \binits{K.T.}},
\bauthor{\bsnm{Knudsen}, \binits{H.A.}},
\bauthor{\bsnm{Ramstad}, \binits{T.}},
\bauthor{\bsnm{L{\o}voll}, \binits{G.}},
\bauthor{\bsnm{M{\aa}l{\o}y}, \binits{K.J.}},
\bauthor{\bsnm{Toussaint}, \binits{R.}},
\bauthor{\bsnm{Flekk{\o}y}, \binits{E.G.}}:
\batitle{Steady-state two-phase flow in porous media: Statistics and transport
  properties}.
\bjtitle{Phys. Rev. Lett.}
\bvolume{102},
\bfpage{074502}
(\byear{2009}).
\doiurl{10.1103/PhysRevLett.102.074502}
\end{barticle}
\endbibitem

\bibitem{rcs11}
\begin{barticle}
\bauthor{\bsnm{Rassi}, \binits{E.M.}},
\bauthor{\bsnm{Codd}, \binits{S.L.}},
\bauthor{\bsnm{Seymour}, \binits{J.D.}}:
\batitle{Nuclear magnetic resonance characterization of the stationary dynamics
  of partially saturated media during steady-state infiltration flow}.
\bjtitle{New J. Phys.}
\bvolume{13},
\bfpage{015007}
(\byear{2011}).
\doiurl{10.1088/1367-2630/13/1/015007}
\end{barticle}
\endbibitem

\bibitem{sbd17}
\begin{barticle}
\bauthor{\bsnm{Sinha}, \binits{S.}},
\bauthor{\bsnm{Bender}, \binits{A.T.}},
\bauthor{\bsnm{Danczyk}, \binits{M.}},
\bauthor{\bsnm{Keepseagle}, \binits{K.}},
\bauthor{\bsnm{Prather}, \binits{C.A.}},
\bauthor{\bsnm{Bray}, \binits{J.M.}},
\bauthor{\bsnm{Thrane}, \binits{L.W.}},
\bauthor{\bsnm{Seymour}, \binits{J.D.}},
\bauthor{\bsnm{Codd}, \binits{S.L.}},
\bauthor{\bsnm{Hansen}, \binits{A.}}:
\batitle{Effective rheology of two-phase flow in three-dimensional porous
  media: experiment and simulation}.
\bjtitle{Transp. Porous Med.}
\bvolume{119},
\bfpage{77}
(\byear{2017}).
\doiurl{10.1007/s11242-017-0874-4}
\end{barticle}
\endbibitem

\bibitem{glb20}
\begin{barticle}
\bauthor{\bsnm{Gao}, \binits{Y.}},
\bauthor{\bsnm{Lin}, \binits{Q.}},
\bauthor{\bsnm{Bijeljic}, \binits{B.}},
\bauthor{\bsnm{Blunt}, \binits{M.J.}}:
\batitle{Pore-scale dynamics and the multiphase darcy law}.
\bjtitle{Phys. Rev. Fluids.}
\bvolume{5},
\bfpage{013801}
(\byear{2020}).
\doiurl{10.1103/PhysRevFluids.5.013801}
\end{barticle}
\endbibitem

\bibitem{zbg21}
\begin{barticle}
\bauthor{\bsnm{Zhang}, \binits{Y.}},
\bauthor{\bsnm{Bijeljic}, \binits{B.}},
\bauthor{\bsnm{Gao}, \binits{Y.}},
\bauthor{\bsnm{Lin}, \binits{Q.}},
\bauthor{\bsnm{Blunt}, \binits{M.J.}}:
\batitle{Quantification of non-linear multiphase flow in porous media}.
\bjtitle{Geophys. Res. Lett.}
\bvolume{48},
\bfpage{2020}--\blpage{090477}
(\byear{2021}).
\doiurl{10.1029/2020GL090477}
\end{barticle}
\endbibitem

\bibitem{tlk09}
\begin{barticle}
\bauthor{\bsnm{Tallakstad}, \binits{K.T.}},
\bauthor{\bsnm{L{\o}voll}, \binits{G.}},
\bauthor{\bsnm{Knudsen}, \binits{H.A.}},
\bauthor{\bsnm{Ramstad}, \binits{T.}},
\bauthor{\bsnm{Flekk{\o}y}, \binits{E.G.}},
\bauthor{\bsnm{M{\aa}l{\o}y}, \binits{K.J.}}:
\batitle{Steady-state, simultaneous two-phase flow in porous media: An
  experimental study}.
\bjtitle{Phys. Rev. E.}
\bvolume{80},
\bfpage{036308}
(\byear{2009}).
\doiurl{10.1103/PhysRevE.80.036308}
\end{barticle}
\endbibitem

\bibitem{sh12}
\begin{barticle}
\bauthor{\bsnm{Sinha}, \binits{S.}},
\bauthor{\bsnm{Hansen}, \binits{A.}}:
\batitle{Effective rheology of immiscible two-phase flow in porous media}.
\bjtitle{Europhys. Lett.}
\bvolume{99},
\bfpage{44004}
(\byear{2012}).
\doiurl{10.1209/0295-5075/99/44004}
\end{barticle}
\endbibitem

\bibitem{yts13}
\begin{barticle}
\bauthor{\bsnm{Yiotis}, \binits{A.G.}},
\bauthor{\bsnm{Talon}, \binits{L.}},
\bauthor{\bsnm{Salin}, \binits{D.}}:
\batitle{Blob population dynamics during immiscible two-phase flows in
  reconstructed porous media}.
\bjtitle{Phys. Rev. E.}
\bvolume{87},
\bfpage{033001}
(\byear{2013}).
\doiurl{10.1103/PhysRevE.87.033001}
\end{barticle}
\endbibitem

\bibitem{sgv21}
\begin{barticle}
\bauthor{\bsnm{Sinha}, \binits{S.}},
\bauthor{\bsnm{Gjennestad}, \binits{M.A.}},
\bauthor{\bsnm{Vassvik}, \binits{M.}},
\bauthor{\bsnm{Hansen}, \binits{A.}}:
\batitle{Fluid meniscus algorithms for dynamic pore-network modeling of
  immiscible two-phase flow in porous media}.
\bjtitle{Front. Phys.}
\bvolume{9},
\bfpage{548497}
(\byear{2021}).
\doiurl{10.3389/fphy.2020.548497}
\end{barticle}
\endbibitem

\bibitem{rh87}
\begin{barticle}
\bauthor{\bsnm{Roux}, \binits{S.}},
\bauthor{\bsnm{Herrmann}, \binits{H.J.}}:
\batitle{Disorder-induced nonlinear conductivity}.
\bjtitle{Europhys. Lett.}
\bvolume{1227},
\bfpage{4}
(\byear{1987}).
\doiurl{10.1209/0295-5075/4/11/003}
\end{barticle}
\endbibitem

\bibitem{rsh21}
\begin{barticle}
\bauthor{\bsnm{Roy}, \binits{S.}},
\bauthor{\bsnm{Sinha}, \binits{S.}},
\bauthor{\bsnm{Hansen}, \binits{A.}}:
\batitle{Role of pore-size distribution on effective rheology of two-phase flow
  in porous media}.
\bjtitle{Front. Water}
\bvolume{3},
\bfpage{709833}
(\byear{2021}).
\doiurl{10.3389/frwa.2021.709833}
\end{barticle}
\endbibitem

\bibitem{fsr21}
\begin{barticle}
\bauthor{\bsnm{Fyhn}, \binits{H.}},
\bauthor{\bsnm{Sinha}, \binits{S.}},
\bauthor{\bsnm{Roy}, \binits{S.}},
\bauthor{\bsnm{Hansen}, \binits{A.}}:
\batitle{Rheology of immiscible two-phase flow in mixed wet porous media:
  Dynamic pore network model and capillary fiber bundle model results}.
\bjtitle{Transp. Porous Med.}
\bvolume{139},
\bfpage{491}
(\byear{2021}).
\doiurl{10.1007/s11242-021-01674-3}
\end{barticle}
\endbibitem

\bibitem{l41}
\begin{barticle}
\bauthor{\bsnm{Leverett}, \binits{M.C.}}:
\batitle{Capillary behavior in porous solids}.
\bjtitle{Trans. AIME}
\bvolume{142},
\bfpage{152}
(\byear{1941}).
\doiurl{10.2118/941152-G}
\end{barticle}
\endbibitem

\bibitem{ly94}
\begin{barticle}
\bauthor{\bsnm{Li}, \binits{X.}},
\bauthor{\bsnm{Yortsos}, \binits{Y.C.}}:
\batitle{Bubble growth and stability in an effective porous medium}.
\bjtitle{Phys. Fluids}
\bvolume{6},
\bfpage{1663}
(\byear{1994}).
\doiurl{10.1063/1.868229}
\end{barticle}
\endbibitem

\bibitem{rk15}
\begin{barticle}
\bauthor{\bsnm{Reynolds}, \binits{C.A.}},
\bauthor{\bsnm{Krevor}, \binits{S.}}:
\batitle{Characterizing flow behavior for gas injection: Relative permeability
  of co2-brine and n2-water in heterogeneous rocks}.
\bjtitle{Water Resources Res.}
\bvolume{51},
\bfpage{9464}
(\byear{2015}).
\doiurl{10.1002/2015WR018046}
\end{barticle}
\endbibitem

\bibitem{lkd15}
\begin{barticle}
\bauthor{\bsnm{Abidoye}, \binits{L.K.}},
\bauthor{\bsnm{Khudaida}, \binits{K.J.}},
\bauthor{\bsnm{Das}, \binits{D.B.}}:
\batitle{Geological carbon sequestration in the context of two-phase flow in
  porous media: A review}.
\bjtitle{Crit. Rev. Env. Sc. Tech.}
\bvolume{45},
\bfpage{1105}
(\byear{2015}).
\doiurl{10.1080/10643389.2014.924184}
\end{barticle}
\endbibitem

\bibitem{ipr19}
\begin{barticle}
\bauthor{\bsnm{Iglauer}, \binits{S.}},
\bauthor{\bsnm{Paluszny}, \binits{A.}},
\bauthor{\bsnm{Rahman}, \binits{T.}},
\bauthor{\bsnm{Zhang}, \binits{Y.}},
\bauthor{\bsnm{W\"{u}lling}, \binits{W.}},
\bauthor{\bsnm{Lebedev}, \binits{M.}}:
\batitle{Residual trapping of co2 in an oil-filled, oil-wet sandstone core:
  Results of three-phase pore-scale imaging}.
\bjtitle{Geophys. Res. Lett.}
\bvolume{46},
\bfpage{11146}
(\byear{2019}).
\doiurl{10.1029/2019GL083401}
\end{barticle}
\endbibitem

\bibitem{nmn20}
\begin{barticle}
\bauthor{\bsnm{Niblett}, \binits{D.}},
\bauthor{\bsnm{Mularczyk}, \binits{A.}},
\bauthor{\bsnm{Niasar}, \binits{V.}},
\bauthor{\bsnm{Eller}, \binits{J.}},
\bauthor{\bsnm{Holmes}, \binits{S.}}:
\batitle{Two-phase flow dynamics in a gas diffusion layer - gas channel -
  microporous layer system}.
\bjtitle{J. Power Sources}
\bvolume{471},
\bfpage{228427}
(\byear{2020}).
\doiurl{10.1016/j.jpowsour.2020.228427}
\end{barticle}
\endbibitem

\bibitem{hzl17}
\begin{barticle}
\bauthor{\bsnm{Huang}, \binits{G.}},
\bauthor{\bsnm{Zhu}, \binits{Y.}},
\bauthor{\bsnm{Liao}, \binits{Z.}},
\bauthor{\bsnm{Ouyang}, \binits{X.L.}},
\bauthor{\bsnm{Jiang}, \binits{P.X.}}:
\batitle{Experimental investigation of transpiration cooling with phase change
  for sintered porous plates}.
\bjtitle{Int. J. Heat Mass Transfer}
\bvolume{114},
\bfpage{1201}
(\byear{2017}).
\doiurl{10.1016/j.ijheatmasstransfer.2017.05.114}
\end{barticle}
\endbibitem

\bibitem{gzk11}
\begin{barticle}
\bauthor{\bsnm{Gedupudi}, \binits{S.}},
\bauthor{\bsnm{Zu}, \binits{Y.Q.}},
\bauthor{\bsnm{Karayiannis}, \binits{T.G.}},
\bauthor{\bsnm{Kenning}, \binits{D.B.R.}},
\bauthor{\bsnm{Yan}, \binits{Y.Y.}}:
\batitle{Confined bubble growth during flow boiling in a mini/micro-channel of
  rectangular cross-section part i: Experiments and 1-d modelling}.
\bjtitle{Inr. J. Therm. Sci.}
\bvolume{50},
\bfpage{250}
(\byear{2011}).
\doiurl{10.1016/j.ijthermalsci.2010.09.001}
\end{barticle}
\endbibitem

\bibitem{lww12}
\begin{barticle}
\bauthor{\bsnm{Li}, \binits{D.}},
\bauthor{\bsnm{Wu}, \binits{G.S.}},
\bauthor{\bsnm{Wang}, \binits{W.}},
\bauthor{\bsnm{Wang}, \binits{Y.D.}},
\bauthor{\bsnm{Liu}, \binits{D.}},
\bauthor{\bsnm{Zhang}, \binits{D.C.}},
\bauthor{\bsnm{Chen}, \binits{Y.F.}},
\bauthor{\bsnm{Peterson}, \binits{G.P.}},
\bauthor{\bsnm{Yang}, \binits{R.}}:
\batitle{Enhancing flow boiling heat transfer in microchannels for thermal
  management with monolithically-integrated silicon nanowires}.
\bjtitle{Nano Lett.}
\bvolume{12},
\bfpage{3385}
(\byear{2012}).
\doiurl{10.1021/nl300049f}
\end{barticle}
\endbibitem

\bibitem{lwy20}
\begin{barticle}
\bauthor{\bsnm{Li}, \binits{W.}},
\bauthor{\bsnm{Wang}, \binits{Z.}},
\bauthor{\bsnm{Yang}, \binits{F.}},
\bauthor{\bsnm{Alam}, \binits{T.}},
\bauthor{\bsnm{Jiang}, \binits{M.}},
\bauthor{\bsnm{Qu}, \binits{X.}},
\bauthor{\bsnm{Kong}, \binits{F.}},
\bauthor{\bsnm{Khan}, \binits{A.S.}},
\bauthor{\bsnm{Liu}, \binits{M.}},
\bauthor{\bsnm{Alwazzan}, \binits{M.}},
\bauthor{\bsnm{Tong}, \binits{Y.}},
\bauthor{\bsnm{Li}, \binits{C.}}:
\batitle{Supercapillary architecture-activated two-phase boundary layer
  structures for highly stable and efficient flow boiling heat transfer}.
\bjtitle{Adv. Matter}
\bvolume{32},
\bfpage{1905117}
(\byear{2020}).
\doiurl{10.1002/adma.201905117}
\end{barticle}
\endbibitem

\bibitem{bs19}
\begin{barticle}
\bauthor{\bsnm{Bremer}, \binits{J.}},
\bauthor{\bsnm{Sundmacher}, \binits{K.}}:
\batitle{Operation range extension via hot-spot control for catalytic co2
  methanation reactors}.
\bjtitle{React. Chem. Eng.}
\bvolume{4},
\bfpage{1019}
(\byear{2019}).
\doiurl{10.1039/C9RE00147F}
\end{barticle}
\endbibitem

\bibitem{sgd16}
\begin{barticle}
\bauthor{\bsnm{Sapin}, \binits{P.}},
\bauthor{\bsnm{Gourbil}, \binits{A.}},
\bauthor{\bsnm{Duru}, \binits{P.}},
\bauthor{\bsnm{Fichot}, \binits{F.}},
\bauthor{\bsnm{Prat}, \binits{M.}},
\bauthor{\bsnm{Quintard}, \binits{M.}}:
\batitle{Reflooding with internal boiling of a heating model porous medium with
  mm-scale pores}.
\bjtitle{Int. J. Heat Mass Transfer}
\bvolume{99},
\bfpage{512}
(\byear{2016}).
\doiurl{10.1016/j.ijheatmasstransfer.2016.04.013}
\end{barticle}
\endbibitem

\bibitem{szx11}
\begin{barticle}
\bauthor{\bsnm{Sun}, \binits{Y.}},
\bauthor{\bsnm{Zhang}, \binits{L.}},
\bauthor{\bsnm{Xu}, \binits{H.}},
\bauthor{\bsnm{Zhong}, \binits{X.}}:
\batitle{Subcooled flow boiling heat transfer from microporous surfaces in a
  small channel}.
\bjtitle{Int. J. Therm. Sci.}
\bvolume{50},
\bfpage{881}
(\byear{2011}).
\doiurl{10.1016/j.ijthermalsci.2011.01.019}
\end{barticle}
\endbibitem

\bibitem{rn94}
\begin{barticle}
\bauthor{\bsnm{C.}, \binits{R.}},
\bauthor{\bsnm{Nimmo}, \binits{J.R.}}:
\batitle{Modeling of soil water retention from saturation to oven dryness}.
\bjtitle{Water Resour. Res.}
\bvolume{30},
\bfpage{701}
(\byear{1994}).
\doiurl{10.1029/93WR03238}
\end{barticle}
\endbibitem

\bibitem{shb13}
\begin{barticle}
\bauthor{\bsnm{Sinha}, \binits{S.}},
\bauthor{\bsnm{Hansen}, \binits{A.}},
\bauthor{\bsnm{Bedeaux}, \binits{D.}},
\bauthor{\bsnm{Kjelstrup}, \binits{S.}}:
\batitle{Effective rheology of bubbles moving in a capillary tube}.
\bjtitle{Phys. Rev. E}
\bvolume{87},
\bfpage{025001}
(\byear{2013}).
\doiurl{10.1103/PhysRevE.87.025001}
\end{barticle}
\endbibitem

\bibitem{b01}
\begin{barticle}
\bauthor{\bsnm{Blunt}, \binits{M.J.}}:
\batitle{Flow in porous media pore-network models and multiphase flow}.
\bjtitle{Current Opinion Colloid \& Interface Science}
\bvolume{6},
\bfpage{197}
(\byear{2001}).
\doiurl{10.1016/S1359-0294(01)00084-X}
\end{barticle}
\endbibitem

\bibitem{mt09}
\begin{barticle}
\bauthor{\bsnm{Meakin}, \binits{P.}},
\bauthor{\bsnm{Tartakovsky}, \binits{A.M.}}:
\batitle{Modeling and simulation of pore-scale multiphase fluid flow and
  reactive transport in fractured and porous media}.
\bjtitle{Reviews of Geophysics}
\bvolume{47},
\bfpage{3002}
(\byear{2009}).
\doiurl{10.1029/2008RG000263}
\end{barticle}
\endbibitem

\bibitem{jh12}
\begin{barticle}
\bauthor{\bsnm{Joekar-Niasar}, \binits{V.}},
\bauthor{\bsnm{M.}, \binits{H.S.}}:
\batitle{Analysis of fundamentals of two-phase flow in porous media using
  dynamic pore-network models: A review}.
\bjtitle{Crit. Rev. Environ. Sci. Technol.}
\bvolume{42},
\bfpage{1895}
(\byear{2012}).
\doiurl{10.1080/10643389.2011.574101}
\end{barticle}
\endbibitem

\bibitem{grz91}
\begin{barticle}
\bauthor{\bsnm{Gunstensen}, \binits{A.K.}},
\bauthor{\bsnm{H.}, \binits{R.D.}},
\bauthor{\bsnm{S.}, \binits{Z.}},
\bauthor{\bsnm{G.}, \binits{Z.}}:
\batitle{Lattice boltzmann model of immiscible fluids}.
\bjtitle{Phys. Rev. A}
\bvolume{43},
\bfpage{4320}
(\byear{1991}).
\doiurl{10.1103/PhysRevA.43.4320}
\end{barticle}
\endbibitem

\bibitem{rin12}
\begin{barticle}
\bauthor{\bsnm{Ramstad}, \binits{N.} \bsuffix{T.~Idowu}},
\bauthor{\bsnm{Nardi}, \binits{C.}},
\bauthor{\bsnm{{\O}ren}, \binits{P.E.}}:
\batitle{Relative permeability calculations from two-phase flow simulations
  directly on digital images of porous rocks}.
\bjtitle{Transp. Porous Media}
\bvolume{94},
\bfpage{487}
(\byear{2012}).
\doiurl{10.1007/s11242-011-9877-8}
\end{barticle}
\endbibitem

\bibitem{qyz17}
\begin{barticle}
\bauthor{\bsnm{Qiu}, \binits{R.F.}},
\bauthor{\bsnm{You}, \binits{Y.C.}},
\bauthor{\bsnm{Zhu}, \binits{C.X.}},
\bauthor{\bsnm{Chen}, \binits{R.Q.}}:
\batitle{Lattice boltzmann simulation for high-speed compressible viscous flows
  with a boundary layer}.
\bjtitle{Appl. Math. Model.}
\bvolume{48},
\bfpage{567}
(\byear{2017}).
\doiurl{10.1016/j.apm.2017.03.016}
\end{barticle}
\endbibitem

\bibitem{gfj20}
\begin{barticle}
\bauthor{\bsnm{Guo}, \binits{S.}},
\bauthor{\bsnm{Feng}, \binits{Y.}},
\bauthor{\bsnm{Jacob}, \binits{J.}},
\bauthor{\bsnm{Renard}, \binits{F.}},
\bauthor{\bsnm{Sagaut}, \binits{P.}}:
\batitle{An efficient lattice boltzmann method for compressible aerodynamics on
  d3q19 lattice}.
\bjtitle{J. Comput. Phys.}
\bvolume{418},
\bfpage{109570}
(\byear{2020}).
\doiurl{10.1016/j.jcp.2020.109570}
\end{barticle}
\endbibitem

\bibitem{w21}
\begin{barticle}
\bauthor{\bsnm{Washburn}, \binits{E.W.}}:
\batitle{The dynamics of capillary flow}.
\bjtitle{Phys. Rev.}
\bvolume{17},
\bfpage{273}
(\byear{1921}).
\doiurl{10.1103/PhysRev.17.273}
\end{barticle}
\endbibitem

\bibitem{vls10}
\begin{barticle}
\bauthor{\bsnm{Vazquez}, \binits{A.}},
\bauthor{\bsnm{Leifer}, \binits{I.}},
\bauthor{\bsnm{S{\'a}nchez}, \binits{R.M.}}:
\batitle{Consideration of the dynamic forces during bubble growth in a
  capillary tube}.
\bjtitle{Chem. Eng. Sc.}
\bvolume{65},
\bfpage{4046}
(\byear{2010}).
\doiurl{10.1016/j.ces.2010.03.041}
\end{barticle}
\endbibitem

\bibitem{w98}
\begin{barticle}
\bauthor{\bsnm{Welch}, \binits{S.W.J.}}:
\batitle{Direct simulation of vapor bubble growth}.
\bjtitle{Int. J. Heat Mass Trans.}
\bvolume{41},
\bfpage{1655}
(\byear{1998}).
\doiurl{10.1016/S0017-9310(97)00285-8}
\end{barticle}
\endbibitem

\bibitem{kwd06}
\begin{barticle}
\bauthor{\bsnm{Kenning}, \binits{D.B.R.}},
\bauthor{\bsnm{Wen}, \binits{D.S.}},
\bauthor{\bsnm{Das}, \binits{K.S.}},
\bauthor{\bsnm{Wilson}, \binits{S.K.}}:
\batitle{Confined growth of a vapour bubble in a capillary tube at initially
  uniform superheat: experiments and modelling}.
\bjtitle{Int. J. Heat Mass Trans.}
\bvolume{49},
\bfpage{4653}
(\byear{2006}).
\doiurl{10.1016/j.ijheatmasstransfer.2006.04.010}
\end{barticle}
\endbibitem

\bibitem{rhs19}
\begin{barticle}
\bauthor{\bsnm{Roy}, \binits{S.}},
\bauthor{\bsnm{Hansen}, \binits{A.}},
\bauthor{\bsnm{Sinha}, \binits{S.}}:
\batitle{Effective rheology of two-phase flow in a capillary fiber bundle
  model}.
\bjtitle{Front. Phys.}
\bvolume{7},
\bfpage{92}
(\byear{2019}).
\doiurl{10.3389/fphy.2019.00092}
\end{barticle}
\endbibitem

\end{thebibliography}

\end{document}